\documentstyle[multicol,aps,pre,epsfig]{revtex}
\newcommand{\be}{\begin{equation}}
\newcommand{\ee}{\end{equation}}
\newcommand{\ba}{\begin{eqnarray}}
\newcommand{\ea}{\end{eqnarray}}

\newcommand{\bnabla}{\mbox{\boldmath $\nabla$}}

\newcommand{\bPi}{\mbox{\boldmath $\Pi$}}

\newcommand{\br}{{\bf r}}
\newcommand{\bk}{{\bf k}}

\newcommand{\bv}{{\bf v}}
\newcommand{\bw}{{\bf w}}
\newcommand{\eps}{\epsilon}

\begin{document}
\draft
\title{Cahn-Hilliard theory for unstable granular flows}
\author{
T.P.C. van Noije
and M.H. Ernst}
\address{Instituut voor Theoretische Fysica, Universiteit Utrecht,
Postbus 80006,
3508 TA Utrecht, The Netherlands}
\maketitle

\begin{abstract}
A Cahn-Hilliard-type theory for hydrodynamic fluctuations
is proposed that gives a quantitative description
of the slowly evolving spatial correlations
and structures in density and flow fields in the early stages
of evolution of freely cooling granular fluids.
Two mechanisms for pattern selection and structure formation are
identified: unstable modes leading to density clustering
(compare spinodal decomposition), and
selective noise reduction (compare peneplanation
in structural geology)
leading to vortex structures.
As time increases, the structure factor
for the density field develops a maximum, which shifts to
smaller wave numbers.
This corresponds to an approximately diffusively growing 
length scale for density clusters.
The spatial velocity 
correlations exhibit algebraic decay
$\sim r^{-d}$ on intermediate length scales.
The theoretical predictions 
for spatial correlation functions and structure factors
agree well with molecular dynamics
simulations of a system of inelastic hard disks.
\end{abstract}


\begin{multicols}{2}
 
\section{Introduction}
\label{sec:2intro}
Granular fluids in rapid flow stand out as an interesting and
complex many-body problem, on which an impressive amount of
experimental, simulation and theoretical results have been gathered
in recent years.
Advanced high speed measuring techniques, used in real experiments,
as well as sophisticated visualization methods, used in computer
experiments, have produced a lot of high quality data,
which have led to intuitive and qualitative explanations.
However, the experimental conditions are either too complex or the
relevant physical parameters unknown to allow quantitative
theoretical modeling and analysis.

On the other hand there exists a large body of analytic results
obtained from hydrodynamic equations, from kinetic theory and from
simplified mathematical models (on instabilities in flows, on
clustering, on transport properties and non-Gaussian features in
velocity distributions) that are not (yet) accessible to
measurements in laboratory
or computer experiments.
There are a few notable exceptions, which will be discussed below
in so far as they are relevant for the subject of the present
paper.

Recently we have proposed a new mesoscopic theory, based on
fluctuating hydrodynamics with unstable modes (Cahn-Hilliard-type
theory) that provides detailed predictions on spatial correlation
functions and structure factors, which permit a {\em quantitative}
comparison with molecular dynamics (MD) simulations on fluids of
inelastic hard spheres.
Preliminary accounts of the results were published in Refs.\
\cite{noije-prl,noije-prerc}.
This paper \cite{thesis} gives a comprehensive account of the underlying
theoretical ideas, quantitative predictions and 
the detailed calculations.
A subsequent paper \cite{orza+brito+ernst} gives a comprehensive
account of the quantitative comparisons with MD simulations,
qualitative explanations where detailed theory is lacking, as
well as
discussions on ranges of validity and unresolved issues.

The inelasticity of the collisions between grains makes driven and
undriven granular fluids behave very differently from 
atomic or molecular fluids.
A dramatic difference with elastic fluids is that 
granular fluids lose kinetic energy through inelastic
collisions and cool if energy is not supplied externally.
In a thermodynamic sense, granular fluids should be considered 
as `open' systems with an energy sink, created by the inelastic
collisions.
This collisional dissipation mechanism introduces
several new time and length scales,
which are often related to instabilities.
In this paper we will focus on spatial correlation functions and
structure factors, and on the underlying instabilities 
in freely
evolving undriven
granular fluids.
There
are two different mechanisms for pattern
formation, one comparable to spinodal decomposition
\cite{langer}, a second one
comparable to peneplanation in structural geology \cite{geo}.
Moreover, we will discuss the importance of different
intermediate scales, related to 
viscosity, heat conductivity and
compressibility, which are controlled by the inelasticity.

\begin{center}
\begin{figure}[h]
\vspace{4cm}
\centerline{\hspace{-8cm}\psfig{figure=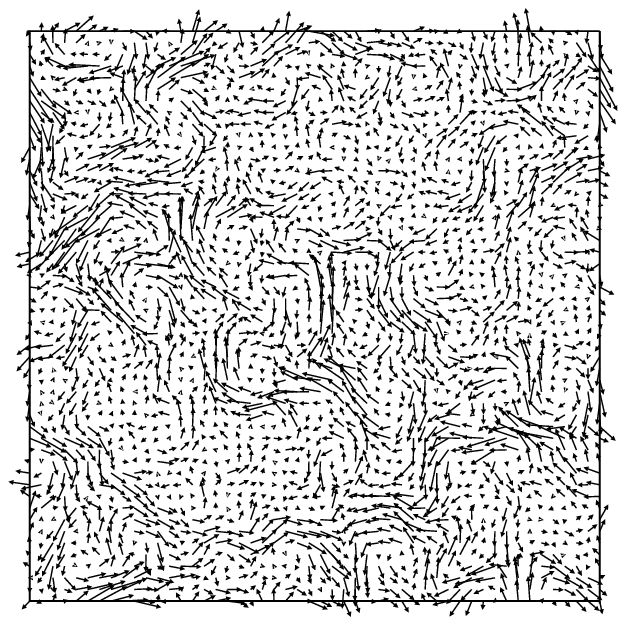,height=8cm}}
\vspace{-7.5cm}
\centerline{\hspace{8cm}\psfig{figure=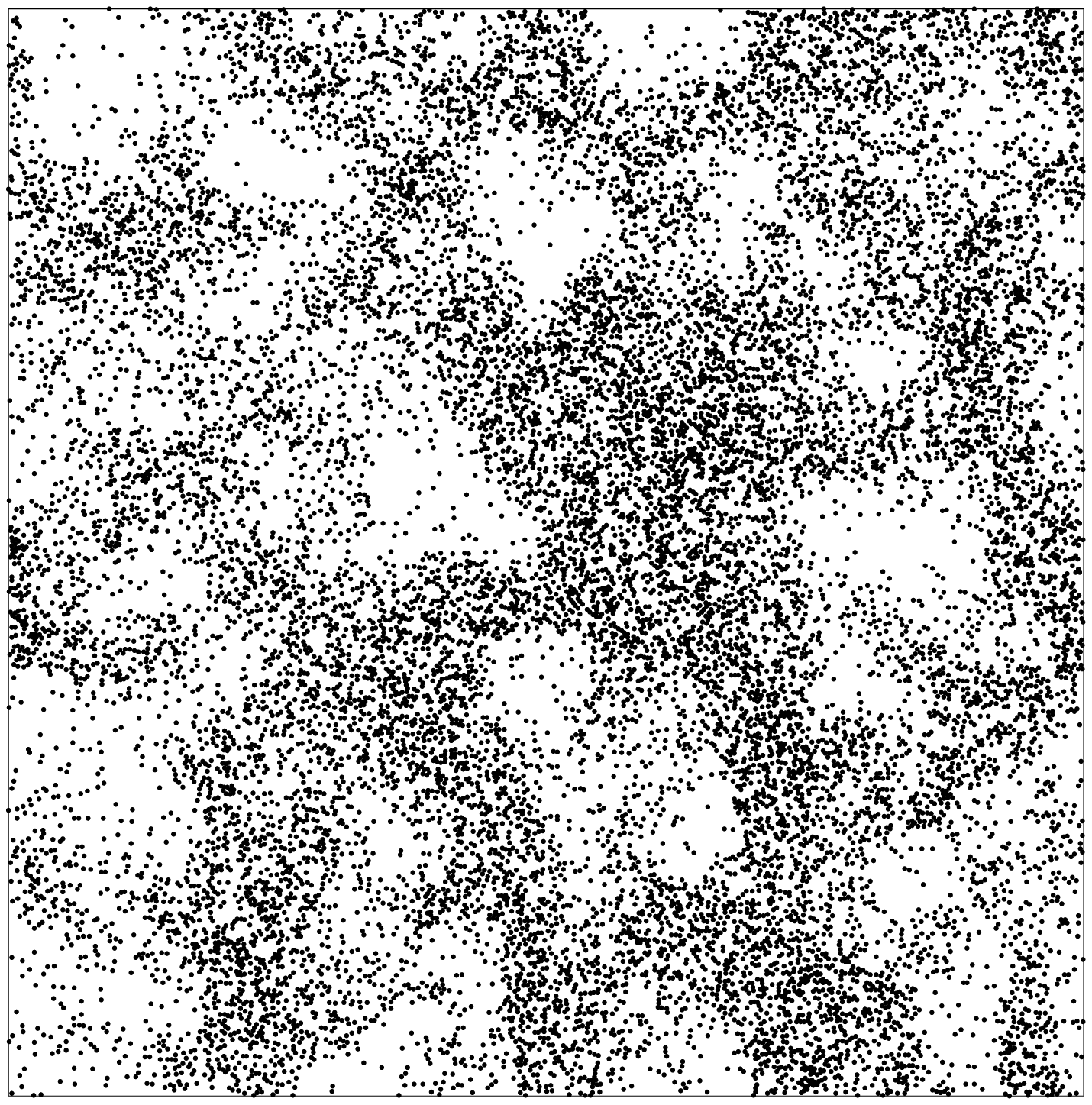,height=7.3cm}}
\vspace{4cm}
\caption{\label{fig:patterns}}
\end{figure}
\end{center}

The instabilities in undriven granular fluids
have been studied 
by several authors 
\cite{goldhirsch,gtz,mcnamara,McNY,brey1,deltour,noije-prl,noije-prerc,brey4,brito+ernst,luding+herrmann}.
Goldhirsch and Zanetti \cite{goldhirsch}
were the first to perform molecular
dynamics (MD) simulations of an undriven two-dimensional (2D)
system of smooth 
inelastic hard disks and
observed the spontaneous formation of density
clusters.
The system is unstable against spatial density
fluctuations, so inhomogeneities in the density field
({\em clusters}) 
slowly grow to macroscopic size.
However, before this happens,
the granular fluid, prepared in a
spatially homogeneous state, remains in a spatially {\em
homogeneous cooling state} (HCS) with a slowly decreasing
temperature. 
Gradually spatial inhomogeneities appear in
the flow field ({\em vortex patterns}), 
and only much later density clusters are being observed.
In Fig.\ \ref{fig:patterns} we show
typical snapshots of the momentum field and the density field, as
obtained in MD simulations of a system of inelastic hard
disks \cite{noije-prl,noije-prerc}.
The flow field develops a large vorticity component and evolves
into a `dense fluid
of closely packed vortex structures', which is still homogeneous
on scales large compared to the average vortex diameter $L_v$,
provided the systems are sufficiently large.
More detailed information on the dynamics of clusters is
available from very recent MD simulations \cite{luding+herrmann}.

The linear stability of undriven granular fluids was first
analyzed by McNamara \cite{mcnamara} on the basis of
hydrodynamic equations for granular fluids.
In Sec.\ \ref{sec:hydrostab} a detailed stability analysis
is presented, which will be needed in analytic calculations of
structure factors and correlation functions. 
The patterns that 
develop in the flow field correspond in fact to a relative
instability: as momentum is conserved in collisions, 
perturbations in the flow field of large enough
wavelength decay slower than the
typical `thermal' velocity, which decays as a result of the
inelasticity of the collisions.
In other words, as the system evolves, the noise from
the thermal motion is reduced, which makes the initial 
long wavelength fluctuations in the flow field observable.
However, their magnitude remains bounded by their values in the
initial homogeneous state.
In $d$-dimensional granular fluids, 
the modes corresponding to this relative instability are the
$(d-1)$ transverse velocity or shear modes, as
well as the long wavelength longitudinal velocity modes.
In Secs.\ \ref{sec:hydrostab} and \ref{sec:meso} we will show how the 
velocity fluctuations, at finite wave numbers, couple to
density fluctuations, resulting 
in a linear instability of the density field.
Unlike the flow field instability, the density
instability is an absolute instability, giving rise to
macroscopic clustering.

\begin{center}
\begin{figure}[h]
\epsfig{figure=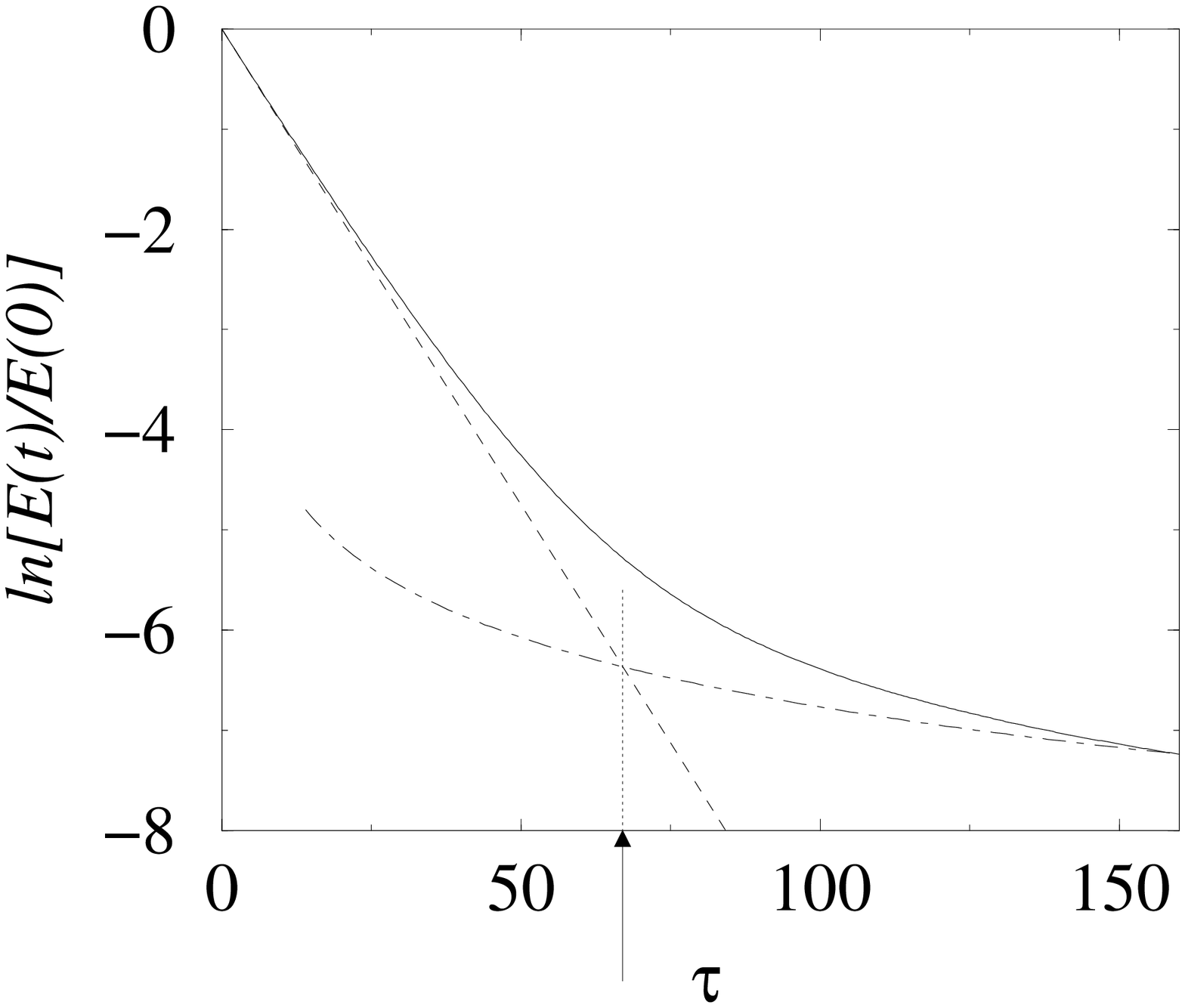,width=10cm,angle=0}
\vspace{1cm}
\caption{\label{fig:cooling}}
\end{figure}
\end{center}

Comparison of Figs.\ \ref{fig:patterns} and \ref{fig:cooling} shows 
how the spatial
inhomogeneities in density and flow field, slow down the cooling.
Initially the system follows Haff's cooling law \cite{haff}, 
as will be briefly discussed in the next section.
Once the inhomogeneities have become important, the homogeneous
cooling law breaks down at a time $\tau_c$, estimated in Fig.\
\ref{fig:cooling} and in Ref.\ \cite{brito+ernst}, 
and the system crosses over to the nonlinear
clustering regime with a slower energy decay.
This happens the more rapidly, the larger the
inelasticity $\epsilon$.
In Ref.\ \cite{brito+ernst}, a
mode coupling theory is described,
which perturbatively takes the effect of
the inhomogeneities on the energy decay into account and
employs the decay of long wavelength fluctuations.
This theory gives quantitative explanations of the energy decay
observed in Fig.\ \ref{fig:cooling} at short {\em and} long
times provided that the inelasticity ($\alpha\gtrsim 0.6$) and
packing fraction ($\phi\lesssim 0.4$ in 2D) are not too large.
A more general mode coupling theory is developed in Ref.\
\cite{pre1}.

In order to understand and quantify the patterns of 
Fig.\ \ref{fig:patterns},
we study the time dependent spatial correlation functions of the
corresponding fluctuations, using Landau and Lifshitz's theory of
fluctuating hydrodynamics
\cite{landau}, adapted to dissipative hard sphere fluids, i.e.\
adapted to the presence of unstable modes.
We present a theory that describes the
buildup of spatial correlations in the density and flow field in
the initial regime, where the inhomogeneities are governed by linear
hydrodynamics, i.e.\ linearized around the homogeneous cooling
state (HCS).
To motivate and explain the theory we compare a freely evolving
granular fluid with spinodal decomposition \cite{langer}, where
the observed phenomena are {\em similar} in several respects.

In spinodal decomposition a mixture is prepared in a spatially
homogeneous state (analogous to the homogeneous cooling state)
by a deep temperature quench into the unstable region.
In this region long wavelength composition fluctuations are
unstable (analogous to density fluctuations in granular fluids)
and lead to phase separation and pattern formation (analogous to
the formation of density clusters and vortex patterns).
This instability and the concomitant pattern formation are
explained by the Cahn-Hilliard (CH) theory \cite{langer}, where
the time dependent structure factor for composition
fluctuations, $S_{nn}(k,t)$, is calculated on the basis of a
macroscopic diffusion equation.
In the CH theory the unstable composition fluctuations
are described by a $k$-dependent diffusion coefficient, which is
{\em negative} below a certain threshold wave number.

The Cahn-Hilliard theory has also been used in the theory of
two-dimensional turbulence (see Frisch \cite{frisch} and
references therein), where the behavior of the fluctuations in
the flow field of incompressible fluids has been described in
terms of negative eddy viscosities.
Vorticity modes with negative effective viscosities have also
been used by Rothman to study vortex formation in lattice gas
cellular automata \cite{rothman}.

The instability of undriven granular fluids  
also {\em differs} in many details from 
spinodal decomposition.
The former is a {\em slow}, 
the latter a {\em fast} process.
Consequently, the Cahn-Hilliard
theory in spinodal decomposition
only describes the onset length and time scales of phase
separation.
As the formation of vortices and clusters in undriven granular
fluids is a rather slow process, the present
theory is expected to give
a good description up to times which are rather large
(see Figs.\ \ref{fig:patterns} and \ref{fig:cooling}) on time
scales that can be reached in molecular dynamics simulations,
provided the inelasticity and the density are not too large.

The dynamics of {\em fluctuations} in granular fluids, say, in
density, $\delta n (\br,t)=n(\br,t)-\langle n \rangle$, 
and flow field, ${\bf
u}(\br,t)$, have hardly been studied
\cite{goldhirsch,deltour,brey4}, 
in sharp contrast to the
average behavior, as witnessed by a
large number of publications. 
We first recall that fluctuations are absent in standard
hydrodynamic as well as
in
Boltzmann-Enskog-type kinetic equations which are based on
molecular chaos (mean field assumption).

The objects of interest in this article are the equal-time
spatial correlation functions of the fluctuating hydrodynamic
fields $\delta a(\br,t)=a(\br,t)-\langle a\rangle$, where $a,b=
\{n,T,u_\alpha\}$ with $\alpha=x,y,\dots$ denoting Cartesian
components,
\be
G_{ab}(\br,t)=V^{-1} \int {\rm d}{\bf r}^\prime \langle
\delta a({\bf r}+\br^\prime,t) \delta b({\bf
r}^\prime,t)\rangle.
\ee
The structure factors are the corresponding Fourier
transforms,
\ba
S_{ab}(\bk,t)&=&\int {\rm d}{\bf r} \exp(-i\bk\cdot \br)
G_{ab}(\br,t)\nonumber\\
&=&V^{-1}\langle \delta a(\bk,t) \delta b(-\bk,t)\rangle,
\label{eq:sdef}
\ea
where $\delta a(\bk,t)$ is the Fourier transform of
$\delta a(\br,t)$, and $V=L^d$ is the volume of the system.
Goldhirsch et al.\ \cite{goldhirsch} initiated MD
studies of $S_{nn}(\bk,t)$ and $S_{pp}(\bk,t)=\sum_\alpha
S_{\alpha\alpha}(\bk,t)$, and related in a qualitative way the
structure at small $k$ to the most unstable shear modes.
They
presented a nonlinear analysis to explain the enslaving of
density fluctuations by the vorticity field.
This analysis reveals that the length scale
associated with the late stages of nonlinear clustering is of the order
$\xi_\perp \sim l_0/\sqrt{\eps}$, where $\eps=1-\alpha^2$
measures the inelasticity and 
$l_0$ is the
mean free path.
Brey et al.\ have also studied the nonlinear response of the
density field to an initially excited $\bk$ mode in the
transverse flow field \cite{brey-nl}.

The most important function to describe the clustering
instability is the structure factor $S_{nn}(k,t)$.
A first step in its theoretical understanding
has been
given by Deltour and Barrat \cite{deltour}.
These authors have shown how
its {\em growth rate} in the linear regime
is determined by the most unstable long wavelength
part of
the heat mode, in which the density couples to longitudinal
velocity perturbations. 

The main goal of the present paper is to study the velocity
and density correlation functions in undriven
granular flows and to demonstrate the importance of internal
fluctuations.
In Secs.\ \ref{sec:meso} and \ref{sec:devo} 
we discuss two different mechanisms for pattern formation, one
leading to density clusters (spinodal decomposition), and one
leading to vortex structures (peneplanation), and we
we calculate the structure factors and show that 
a mesoscopic description in terms of 
fluctuating hydrodynamics \cite{landau}
gives quantitative predictions for the spatial correlations
functions
over large intermediate time intervals,
controlled by linearized hydrodynamics.

In Sec.\ \ref{sec:spatial2} an
analytic description of the correlation
function $G_{\alpha\beta}(\br,t)$ of the components
$u_\alpha(\br,t)$
of the flow field will be given, based on
fluctuating hydrodynamics and
the assumption of incompressible flow.
This theory was presented in Ref.\ \cite{noije-prl} and 
yields predictions, including long range tails
$\sim r^{-d}$ in $d$-dimensional fluids, that
for nearly elastic particles ($\alpha\gtrsim 0.9$)
agree well with two-dimensional  
MD simulations up to large
distances.
As the transverse velocity fluctuations of an incompressible
fluid
do not couple to the density fluctuations in a linear theory, 
this theory 
gives no information on $S_{nn}(k,t)$.
A quantitative theory for the
magnitude of
$S_{nn}(k,t)$ and its $k$ dependence in the full range of
hydrodynamic
wave numbers was given in Ref.\ \cite{noije-prerc} and is
described in the second part of Sec.\ \ref{sec:spatial2}, where the 
assumption of incompressibility is dropped and the
description 
is extended to the general
compressible case, based on the full set
of fluctuating hydrodynamic equations.
The results are also compared with computer simulations.
We end with some conclusions in Sec.\ \ref{sec:concl2}.

\section{Hydrodynamic stability}
\label{sec:hydrostab}
In the present section we discuss the macroscopic hydrodynamic
equations for inelastic hard sphere (IHS) fluids. 
Fluctuations will be considered in the
next section.
We assume that IHS hydrodynamics for weakly inelastic systems
can be described by the standard hydrodynamic conservation
equations
supplemented by a sink term $\Gamma$ in the 
temperature balance equation,
\ba
&&\partial_t n +  \bnabla\cdot (n {\bf u})=0\nonumber\\
&&\partial_t {\bf u}+ {\bf u}
\cdot
\bnabla {\bf u}= -\frac{1}{\rho}  \bnabla\cdot  \bPi\nonumber\\
&&\partial_t T + {\bf u}\cdot  \bnabla T
= -\frac{2}{d n}  (\bnabla\cdot {\bf J}
+\bPi: \bnabla {\bf u})
-\Gamma.
\label{eq:hydrofree}
\ea
Here $\rho=m n$, ${\bf u}$ the flow velocity, and
$\textstyle{\frac{1}{2}} d nT$ the kinetic energy density in the
local rest frame of the IHS fluid.
The pressure tensor $\Pi_{\alpha\beta}=p \delta_{\alpha\beta}
+\delta\Pi_{\alpha\beta}$ contains the local pressure $p$ and the
dissipative momentum flux
$\delta\Pi_{\alpha\beta}$, which is proportional to
$\nabla_\alpha
u_\beta$ and contains the kinematic and longitudinal viscosities
$\nu$
and $\nu_l$, defined below Eq.\ (\ref{eq:app2_1}) of App.\ 
\ref{sec:app2a}.
The constitutive relation for the heat flux, ${\bf J}=-\kappa
\bnabla
T$, defines the heat conductivity $\kappa$.
The above equations can be justified to lowest order in the
inelasticity
$\eps=1-\alpha^2$ \cite{lun,jenkins+richman}, and will be used
for {\em weakly inelastic} systems, where the pressure
is assumed to be given by its value for {\em elastic} hard spheres,
$p=n T[1+2 b(n)]$ with $b(n)$ given below (\ref{eq:app2_1}) in App.\
\ref{sec:app2a}.
Note that ${\cal O}(\eps)$ contributions are being neglected
here, because the Enskog-Boltzmann equation for IHS fluids predicts
$p=n T[1+(1+\alpha) b(n)]$.
Similarly, 
the transport coefficients $\nu$,
$\nu_l$,
and $\kappa$ are assumed to be given by the Enskog theory for a
dense
gas of {\em elastic} hard spheres (EHS) or disks \cite{chapman}.
In the present paper,
corrections of ${\cal O}(\eps)$ to the leading
nonvanishing behavior of thermodynamic properties and transport
fluxes will be neglected.
More general expressions for transport coefficients have been
derived in Refs.\
\cite{goldshtein,sela,brey3,granmat,greenk}.
They contain an
additional contribution, $-\mu \bnabla n$, in the heat flux,
where the new transport coefficient $\mu\sim \eps$ vanishes for
small inelasticity.

Kinetic theory provides an exact expression for the collisional
dissipation rate $\Gamma$, which represents the average energy
loss through inelastic collisions.
It can be derived from the microscopic energy loss per
collision.
An explicit derivation can be found in Refs.\ 
\cite{haff,goldshtein,granmat}. 
For the present purpose however, a phenomenological derivation
suffices, which proceeds as follows.
On average, a particle loses per collision an amount $\sim \eps
T$ of its kinetic energy, and per unit time an amount $\sim \eps
\omega T$,
where $\omega$ is the average collision frequency
\cite{chapman}.
This argument gives (apart from a numerical factor $1/d$)
\be
\Gamma=2\gamma_0 \omega T,
\label{eq:Gloss}
\ee
where $\gamma_0=(1-\alpha^2)/2d$.
In the present theory the collision frequency is assumed to be
given by 
Enskog's theory for dense hard sphere fluids \cite{chapman}, and quoted
explicitly in Eq.\ (\ref{eq:app2}) of App.\ \ref{sec:app2a}.
It is proportional to $\sqrt{T}$.

For an understanding of what follows 
two properties of undriven granular fluids are important: 
(i) the existence of a
{\em homogeneous cooling state} (HCS) and (ii) its {\em 
linear instability}
against spatial fluctuations.
We first observe that the hydrodynamic equations for an IHS fluid, 
initialized in a homogeneous state with
temperature $T_0$, admit an HCS solution
with a homogeneous density $n$, a flow
field which can be set to zero everywhere, and a homogeneous 
temperature $T(t)$, determined by $\partial_t T=-\Gamma$.
To solve this equation it is convenient to change to a new time
variable, defined as ${\rm d}\tau =\omega (T(t)) {\rm d} t$,
yielding $T(t)=T_0\exp(-2 \gamma_0 \tau)$.
To find the relation between the `internal' time $\tau$, which measures 
the average number of collisions 
suffered per particle within a time $t$, and the `external' time
$t$, we integrate the relation for ${\rm d}\tau$ using
$\omega\sim \sqrt{T}$, with the result
\be
\tau=\frac{1}{\gamma_0}\ln[1+\gamma_0 t/t_0].
\label{eq:tauvst}
\ee
In the elastic limit ($\eps\rightarrow 0$), it is proportional to
the external time,
$\tau=t/t_0$, measured in units of the mean free time 
$t_0=1/\omega(T_0)$ at the initial temperature.
The theoretical prediction (\ref{eq:tauvst})
agrees well with the computer simulations 
\cite{orza+brito+ernst} until the system crosses over at $\tau_c$ into
the nonlinear clustering regime, to be discussed below.
The initial slope of $\tau(t)$ corresponds to the collision
frequency $\omega(T_0)$ in the equilibrium state at $\tau=0$.
The relation (\ref{eq:tauvst})
introduces a new intermediate (mesoscopic) time scale, the
homogeneous cooling time $t_e=t_0/\gamma_0$.
Combination of these results yields
the slow decay of the temperature,
\be
T(t)=T_0\exp(-2 \gamma_0 \tau)=T_0/(1+\gamma_0 
t/t_0)^2,
\label{eq:haff}
\ee
which is Haff's well-known homogeneous cooling law \cite{haff}.
There exist extensive verifications of the validity of Haff's
law in the HCS ($\tau\ll\tau_c$ in Fig.\ \ref{fig:cooling})
using MD simulations \cite{gtz,McNY,deltour}.
We show below that this HCS solution is linearly unstable once the
linear extent $L$ of the system exceeds some dynamic correlation
length $\sim l_0/\sqrt{\eps}$
\cite{goldhirsch,mcnamara},
where $l_0$ is the (time independent) mean free path.
It is given by 
$l_0=v_0/\omega$, where
$v_0=\sqrt{2T/m}$ is the thermal velocity.

In general the HCS is highly nontrivial, as it exhibits 
correlations between the velocities and positions of
different particles.
In the `lowest order' description (for more refined
approximations see
\cite{goldshtein,granmat,brilliantov+poeschel})
the HCS corresponds to an equilibrium state, 
which is cooling adiabatically, i.e.\ with time dependent temperature
(\ref{eq:haff}).
Here velocity correlations
between different particles are absent, and position correlations
are taken into account through the pair correlation function at
contact (see (\ref{eq:app2}) of App.\ \ref{sec:app2a}).

In the present paper, 
we are interested in the buildup of correlations between spatial
fluctuations in a system
that is prepared in a homogeneous state at an initial temperature
$T_0$.
It reaches the HCS
within a few mean free times $t_0$.
Therefore, we can linearize Eqs.\ (\ref{eq:hydrofree})
around the homogeneous density $n$ and temperature
$T(t)=T_0 /[1+\gamma_0 t/t_0]^2$, and the vanishing flow field 
of the slowly evolving HCS.
The resulting set of linearized hydrodynamic equations, given
in (\ref{eq:app2_1}) of App.\ \ref{sec:app2a}, contains the 
Enskog hard sphere
transport coefficients, $\nu$, $\nu_l$ and $\kappa$,
which are proportional to $\sqrt{T(t)}$, and
depend therefore explicitly on time.
It is again convenient to make the transformation 
${\rm d}\tau=\omega {\rm d}t$, and
introduce the dimensionless variables 
$\delta \tilde{n}(\br,\tau)= \delta
n(\br,t)/n$, $\tilde{\bf u}(\br,\tau)= {\bf u}(\br,t)/v_0(t)$,
and $\delta \widetilde{T}(\br,\tau)=\delta T(\br,t)/T(t)$. 
In these new variables the equations of change for the
macroscopic Fourier modes $\delta \tilde{n}(\bk,\tau)$, 
$\tilde{\bf u}(\bk,\tau)$, and $\delta \widetilde{T}(\bk,\tau)$,
defined through $\delta\tilde{a}(\bk,\tau)=\int{\rm d}\br
\exp(-i\bk\cdot\br)\delta\tilde{a}(\br,\tau)$,
become
ordinary differential equations with {\em time independent}
coefficients (valid for $k l_0\lesssim 1$), which can be
analyzed in terms of eigenvalues and eigenvectors,
\ba
\frac{\partial \delta \tilde{n}}{\partial \tau} &=& - i k
l_0 \tilde{u}_l\nonumber\\
\frac{\partial \tilde{u}_{\perp}}{\partial \tau} &=& \gamma_0
(1-k^2
\xi_\perp^2)
\tilde{u}_{\perp}\nonumber\\
\frac{\partial \tilde{u}_l}{\partial \tau}&=&\gamma_0 (1-k^2 \xi_l^2) 
\tilde{u}_l
-i k l_0 \left(\frac{p}{2 n T}\right)
\delta \widetilde{T}\nonumber\\
&&- i k l_0 \left(\frac{1}{2 n T \chi_T}
\right)
\delta \tilde{n}\nonumber\\
\frac{\partial \delta \widetilde{T}}{\partial \tau}&=& -\gamma_0 (1+k^2
\xi_T^2) \delta \widetilde{T} - i k l_0 \left(\frac{2
p}
{ d n
T}\right) \tilde{u}_l\nonumber\\
&&- \gamma_0 g(n)
\delta \tilde{n}.
\label{eq:change}
\ea
Here we have introduced 
the {\em time independent} correlation lengths
$\xi_\perp$, $\xi_l$ and
$\xi_T$, defined by $\xi_\perp^2=\nu/\omega
\gamma_0$,
$\xi_l^2 =\nu_l/\omega\gamma_0$
and $\xi_T^2=2 \kappa/d n \omega \gamma_0$, the isothermal
compressibility $\chi_T= {(\partial
n/\partial p)}_T/n$, and for the function
$g(n)=2[1+(n/\chi)\partial
\chi/\partial n]$ we refer to Eqs.\ (\ref{eq:app2_1}) and
(\ref{eq:app2}) of App.\ \ref{sec:app2a}.
The subscript $\perp$ in the equation for $\tilde{u}_\perp$ refers
to any of the $(d-1)$ directions perpendicular to $\bk$, and the
subscript $l$ denotes the longitudinal direction along $\bk$.
The validity of the hydrodynamic Eqs.\ (\ref{eq:change}) is
restricted to wave numbers $k\ll 2\pi/l_0$ to guarantee {\em
separation} of kinetic and hydrodynamic scales, and to $k\ll
2\pi/\sigma$, where $\sigma$ is the diameter of a disk or sphere,
to guarantee that the Euler equations involve only local
hydrodynamics.
Therefore, the validity of the hydrodynamic Eqs.\ (\ref{eq:change}) are
restricted to long wavelengths, satisfying $k\ll {\rm
min}\{2\pi/l_0,2\pi/\sigma\}$.
In matrix representation we write the above equations as
\be
\frac{\partial}{\partial \tau}\delta \tilde{\bf a}(\bk,\tau) = 
\widetilde{\bf
M}(\bk)\delta \tilde{\bf a}(\bk,\tau),
\label{eq:lin2}
\ee
where components of $\tilde{\bf a}$ are labeled with
$\{n,T,l,\perp\}$,
and the hydrodynamic matrix $\widetilde{\bf M}$ is defined by Eqs.\
(\ref{eq:change}).
The linear stability of (\ref{eq:lin2}) was first investigated
in Refs.\ \cite{goldhirsch,mcnamara}.
Its eigenvalues or dispersion relations $\zeta_\lambda(k)$ and
corresponding eigenvectors are analyzed in App.\
\ref{sec:app2a}.
The $\zeta_\lambda(k)$ were first calculated
numerically
in Ref. \cite{mcnamara}, using transport
coefficients somewhat different 
from those obtained from the Enskog theory.
Typical results of our calculations are shown in Fig.\ \ref{fig:disp}.
The most striking feature is that there are two 
eigenvalues, $\zeta_\perp$ and $\zeta_H$, that are {\em positive}
below the threshold values $k_\perp^\ast$ and $k_H^\ast$,
i.e.\ two linearly
unstable modes with exponential growth rates.
The spectrum has different characteristics for smaller and larger
wavelengths.
In the {\em dissipative} range \cite{mcnamara} ($k l_0 \ll
\gamma_0$) all eigenvalues are real; propagating modes are
absent.
Around $k l_0 \sim {\cal O}(\gamma_0)$, two eigenvalues become
complex conjugates and the corresponding (sound)
modes become propagating.
In the {\em standard} range ($k l_0\gg \gamma_0$), compression effects
and sound propagation, which are ${\cal O}(k l_0)$,
dominate dissipation, whereas heat conduction, which is ${\cal
O}(k^2 l_0^2)$, becomes dominant only
in the elastic range ($k l_0\gg \sqrt{\gamma_0}$), where $\gamma_0$
is assumed to be sufficiently small so that $\gamma_0 \ll
\sqrt{\gamma_0}$.
In the latter range,
the dispersion relations and eigenmodes resemble those of an
elastic fluid.
In App.\ \ref{sec:app2a} the full set of eigenvalues 
$\zeta_\lambda(k)$ and
eigenmodes is analyzed.
Here we only analyze the $(d-1)$-fold
degenerate transverse velocity or shear modes, which are
decoupled from the remaining modes.
In fact, Eqs.\ (\ref{eq:change}) show already that $\zeta_\perp(k)=\gamma_0
(1-k^2\xi_\perp^2)$, which describes an {\em unstable} mode for
$k<k_\perp^\ast=1/\xi_\perp$.

\begin{center}
\begin{figure}[h]
\epsfig{figure=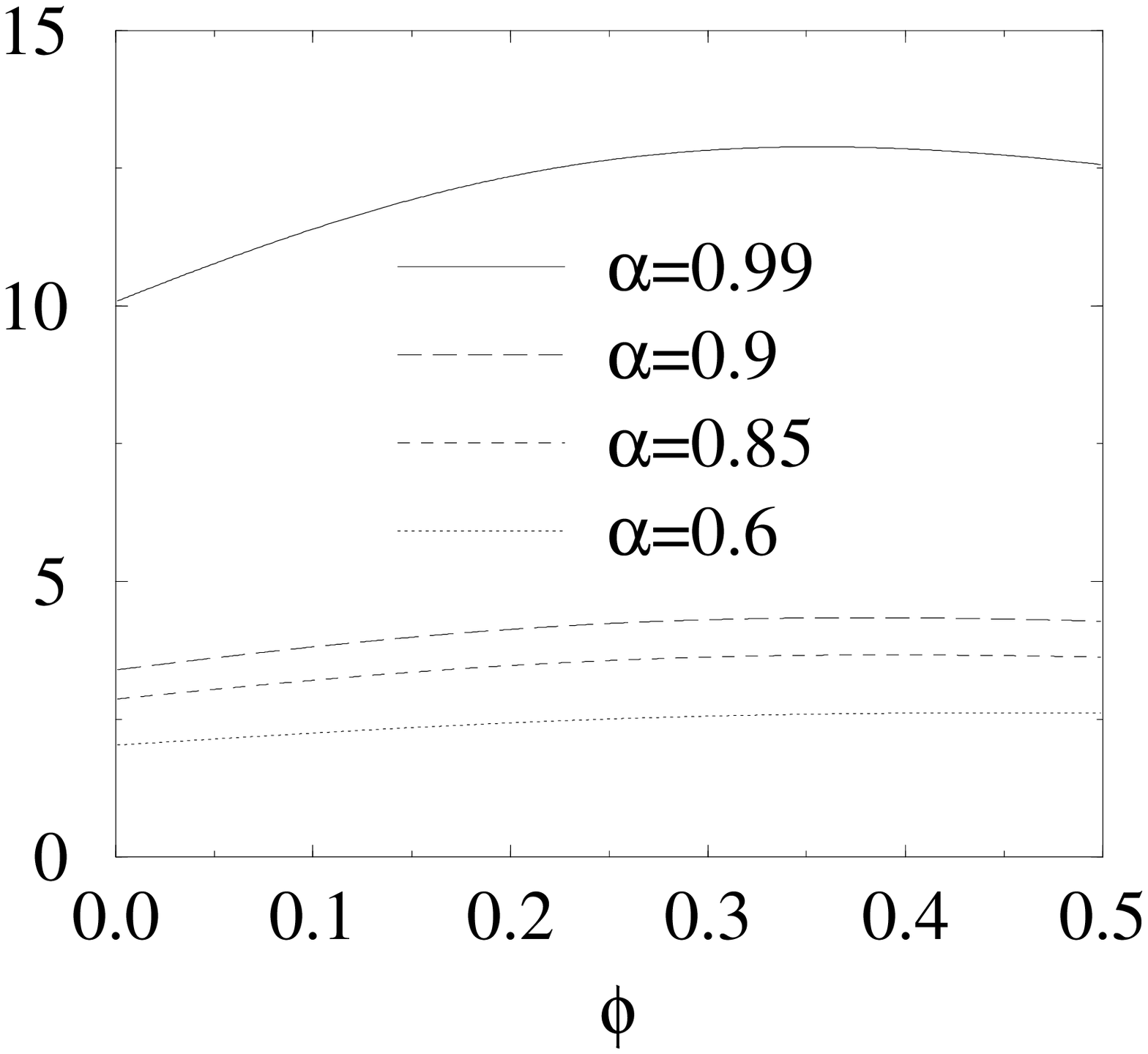,width=10cm,angle=0}
\vspace{2cm}
\caption{\label{fig:ratio}}
\end{figure}
\end{center}

Similarly, the heat mode is unstable for $k<k_H^\ast\propto
1/\xi_T$, as shown in App.\ \ref{sec:app2a}.
In the dissipative range ($k l_0\ll \gamma_0$), the heat mode is
given by $\zeta_H(k)=\gamma_0(1-k^2\xi_\parallel^2)$ where
$\xi_\parallel$ is calculated in App.\ \ref{sec:app2a}.
In this range the heat mode is a pure longitudinal velocity
fluctuation.
We point out that $\xi_\perp$ diverges as
$1/\sqrt{\eps}$ for small $\eps$, while $\xi_\parallel\sim
1/\eps$ [see (\ref{eq:xip})].
As a consequence the correlation lengths $\xi_\perp$ and
$\xi_\parallel$ are well separated for small inelasticity,
as shown in Fig.\ \ref{fig:ratio}. 

Furthermore, we observe that the instability of shear and heat mode 
is a long
wavelength instability. As a consequence, for finite systems
effects of the
boundaries are important,
and the various instabilities are suppressed in small systems.
When using periodic boundary conditions, the instabilities are
suppressed if 
$k_{\rm min}=2\pi/L$ is larger than $k_\perp^\ast$ or
$k_H^\ast$.
When decreasing the system length $L=V^{1/d}$ at fixed inelasticity,
first the heat mode will become {\em stable} ($k_H^\ast<k_{\rm
min} <
k_\perp^\ast$).
In this range
the density (coupled to the heat mode) is {\em linearly} stable,
and
density
inhomogeneities 
can only be created via a nonlinear 
coupling to the
unstable shear mode \cite{goldhirsch}.
Decreasing the
system
size even further ($k_\perp^\ast<k_{\rm min}$) will stabilize the 
shear mode and thus the HCS
itself.
In the next section we present a mesoscopic theory to
describe the dynamics of the long wavelength fluctuations 
in the system.

\section{Mesoscopic hydrodynamics}
\label{sec:meso}
In this section we study spatial fluctuations, around a reference
state, of the hydrodynamic fields, which leads to a
Cahn-Hilliard-type theory \cite{langer} for the structure factors.
The dynamics of these fluctuations can be described by the
fluctuating hydrodynamic equations \cite{landau}, obtained from
the nonlinear equations (\ref{eq:hydrofree}) by adding
fluctuation terms to the momentum and heat current, denoted by
$\hat{\bPi}$ and $\hat{\bf J}$ respectively.
These currents $\hat{\bPi}$ and $\hat{\bf J}$ are considered as
Gaussian white noise, local in space, and their correlations are
determined by some appropriately formulated
fluctuation-dissipation theorem for the reference state.

In elastic fluids the reference state would be the thermal
equilibrium state.
In driven systems it would be a nonequilibrium steady state
(NESS).
In the present case 
the reference state is the slowly evolving homogeneous cooling
state.
In lowest approximation \cite{granmat} it may be considered as an
adiabatically changing equilibrium state with a constant density,
a vanishing flow field and a time dependent temperature,
described by Haff's law (\ref{eq:haff}). 
The basic extension required for application to IHS fluids, is
the assumption that the fluctuation-dissipation theorem also
applies to the HCS with an adiabatically changing temperature
$T(t)$.
This assumption
relates the noise strengths
to the transport coefficients through \cite{landau}
\ba
&&\langle\hat{\Pi}_{\alpha\beta}(\br,t)\hat{\Pi}_{\gamma\delta}(\br^\prime,
t^\prime
)\rangle=2 T[\eta(\delta_{\alpha\gamma}\delta_{\beta\delta}
+\delta_{\alpha\delta}\delta_{\beta\gamma})\nonumber\\
&&\;\;\;\;\;\;\;\;\;\;\;\;\;\;\;\;\;\;
+(\zeta-\frac{2}{d}\eta)\delta_{\alpha\beta}\delta_{\gamma\delta}]
\delta(\br-\br
^\prime)\delta(t-t^\prime)\nonumber\\
&&\langle \hat{J}_\alpha
(\br,t)\hat{J}_\beta(\br^\prime,t^\prime)\rangle=2\kappa T^2
\delta_{\alpha\beta} \delta(\br-\br^\prime)\delta(t-t^\prime).
\label{eq:noi2}
\ea
For dimensional reasons, the transport coefficients in systems
with hard sphere type interactions, like IHS, are
proportional to 
$\sqrt{T(t)}$.

In the present theory for nearly elastic fluids
we linearize the nonlinear Langevin
equations, obtained from (\ref{eq:hydrofree}), around the HCS.
By applying the transformations introduced above Eqs.\
(\ref{eq:change}) we obtain the dynamic equations for the
rescaled variables, in the form of a set of Langevin 
equations
with constant coefficients $\widetilde{\bf M}(\bk)$, i.e.\
\be
\frac{\partial}{\partial \tau}\delta\tilde{\bf a}(\bk,\tau) = 
\widetilde{\bf
M}(\bk)\delta\tilde{\bf a}(\bk,\tau)
+\hat{\bf
f}(\bk,\tau),
\label{eq:match}
\ee
where $\hat{\bf f}$ represents the rescaled internal fluctuations in
the momentum and heat flux.

The equation of motion for the matrix of rescaled structure
factors, $\tilde{S}_{ab}(\bk,\tau)=V^{-1}\langle \delta
\tilde{a}(\bk,\tau)\delta \tilde{b}(-\bk,\tau)\rangle$, can now
be derived from Eq.\ (\ref{eq:match}), and yields
\be
\frac{\partial}{\partial \tau} \tilde{\bf S}(\bk,\tau)
= \widetilde{\bf M}(\bk)\cdot\tilde{\bf S}(\bk,\tau)
+\tilde{\bf S}(\bk,\tau)\cdot\widetilde{\bf M}^T(-\bk)
+  \tilde{\bf C}(k),
\label{eq:matrixfl}
\ee
where $\widetilde{\bf M}^T$ is the transpose of $\widetilde{\bf M}$.
It is to be solved for given initial values $\tilde{\bf S}(\bk,0)$.
In terms of the rescaled variables the Gaussian white noise 
(\ref{eq:noi2}) has
the standard from with {\em constant} coefficients, given by the
covariance matrix, $\tilde{\bf C}(k)$, with
\be
V^{-1}\langle
\hat{f}_a(\bk,\tau)\hat{f}_b
(-\bk,\tau^\prime)\rangle= \tilde{C}_{ab}(k)
\delta(\tau-\tau^\prime).
\label{eq:snoi2}
\ee
It is diagonal with nonvanishing elements
\ba
\tilde{C}_{TT}&=& 4 
\gamma_0 k^2 \xi_T^2/d n \nonumber\\
\tilde{C}_{ll}&=& \gamma_0 k^2 \xi_l^2/n\nonumber\\
\tilde{C}_{\perp\perp}&=& \gamma_0 k^2 \xi_\perp^2/n,
\label{eq:strengthf}
\ea
as follows from (\ref{eq:noi2}) and the definitions of the
correlation lengths below (\ref{eq:change}).
The formal solution of (\ref{eq:matrixfl}) for the matrix of
rescaled structure factors is then 
\ba
\tilde{\bf S}(\bk,\tau)=&\exp[\widetilde{\bf M}(\bk)\tau]\cdot\tilde{\bf
S}(\bk,0)\cdot\exp[\widetilde{\bf M}^T(-\bk)\tau]\nonumber\\
+\int_0^\tau{\rm d}\tau^\prime &\exp[\widetilde{\bf
M}(\bk)\tau^\prime]\cdot\tilde{\bf C}(k)\cdot\exp[\widetilde{\bf
M}^T(-\bk)\tau^\prime].
\label{eq:stab}
\ea
At the initial time ($\tau=0$) the system is prepared in a
thermal equilibrium state 
of elastic hard spheres
with density $n$ and temperature $T_0$.
Consequently, all elements of $\tilde{\bf S}(\bk,0)$ are known.
Moreover, the evolution equations (\ref{eq:matrixfl}) and
(\ref{eq:change}) are only valid for $k\ll \{2\pi/l_0,2\pi/\sigma\}$.
So the initial values $\tilde{\bf S}(\bk,0)$ are only needed for
$k\sigma\ll 2\pi$, where they are given by their limiting values\footnote{Note that $\bk$ should not be set
equal to ${\bf 0}$, because
$\delta \tilde{n}({\bf 0},t)=0$ when the total number of particles
is fixed.}
as $k\rightarrow 0$.
The nonvanishing $\tilde{S}_{ab}(\bk,0)$ with
$a,b=\{n,T,l,\perp\}$ are given by their equipartition
values for elastic hard sphere fluids,
\ba
\tilde{S}_{nn}(k,0)&=&
 T \chi_T\nonumber\\
\tilde{S}_{TT}(k,0)&=&\frac{2}{dn}
\nonumber\\
\tilde{S}_{ll}(k,0)&=&\tilde{S}_{\perp\perp}(k,0)=\frac{1}{2n}.
\label{eq:init}
\ea
The above equations can be solved numerically or analytically.
For a theoretical analysis it is more convenient to study the
deviations from the initial values in {\em thermal equilibrium},
defined as
\be
\tilde{\bf S}^+(\bk,\tau)=\tilde{\bf
S}(\bk,\tau)-\tilde{\bf S}(\bk,0).
\label{eq:dels2}
\ee
The reason is that $\tilde{\bf S}(\bk,\tau)$ itself at {\em larger}
$k$ values also approaches $\tilde{\bf S}(\bk,0)$ by the imposed
fluctuation-dissipation theorem (\ref{eq:noi2}) or
(\ref{eq:snoi2}).
Inspection of (\ref{eq:matrixfl}) for $k l_0 \gg \sqrt{\gamma_0}$
(elastic range) shows that all $k$ {\em independent} terms in the
matrix $\widetilde{\bf M}(\bk)$ in (\ref{eq:change}) can be {\em
neglected} and the hydrodynamic matrix reduces to the elastic
one, $\tilde{\bf E}(\bk)$.
Consequently, Eq.\ (\ref{eq:matrixfl}) reduces to 
\be
\tilde{\bf E}(\bk)\cdot\tilde{\bf S}(\bk,0)
+\tilde{\bf S}(\bk,0)\cdot\tilde{\bf E}^T(-\bk)
+  \tilde{\bf C}(k)={\bf 0},
\label{eq:matrixfle}
\ee
as can be verified using (\ref{eq:strengthf}) and
(\ref{eq:init}).

Subtracting this equation from (\ref{eq:matrixfl}) yields an
equation of a similar form as (\ref{eq:matrixfl}),
\ba
\frac{\partial}{\partial \tau} \tilde{\bf S}^+(\bk,\tau)
&=& \widetilde{\bf M}(\bk)\cdot\tilde{\bf S}^+(\bk,\tau)\nonumber\\
&+&\tilde{\bf S}^+(\bk,\tau)\cdot\widetilde{\bf M}^T(-\bk)
+  \tilde{\bf B}(k),
\label{eq:matrixfld}
\ea
with a source term
\ba
\tilde{\bf B}(k)&=&[\widetilde{\bf M}(\bk)-\tilde{\bf E}(\bk)]\cdot\tilde{\bf
S}(\bk,0)\nonumber\\
&+&\tilde{\bf S}(\bk,0)\cdot[\widetilde{\bf
M}^T(-\bk)-\tilde{\bf E}^T(-\bk)].
\ea
Its nonvanishing matrix elements are
\ba
\tilde{B}_{nT}&=&\tilde{B}_{Tn}=-\gamma_0 g(n) \tilde{S}_{nn}(k,0)\nonumber\\
\tilde{B}_{TT}&=&-2 \gamma_0 \tilde{S}_{TT}(k,0)\nonumber\\
\tilde{B}_{ll}&=&2\gamma_0 \tilde{S}_{ll}(k,0)\nonumber\\
\tilde{B}_{\perp\perp}&=&2\gamma_0
\tilde{S}_{\perp\perp}(k,0),
\label{eq:b2}
\ea
and the formal solution of (\ref{eq:matrixfld}) with the initial
value $\tilde{\bf S}^+(\bk,0)=0$ becomes,
\be
\tilde{\bf S}^+(\bk,\tau)=
\int_0^\tau{\rm d}\tau^\prime \exp[\widetilde{\bf
M}(\bk)\tau^\prime]\cdot\tilde{\bf B}(k)\cdot\exp[\widetilde{\bf
M}^T(-\bk)\tau^\prime].
\label{eq:intds}
\ee
The spectral decomposition (\ref{eq:spde}) of App.\
\ref{sec:app2a} allows us to write the rescaled structure factors
for $a,b=\{n,T,l,\perp\}$ as
\ba
\tilde{S}^+_{ab}(\bk,\tau)&=&\sum_{\lambda\mu} \tilde{w}_{\lambda
a}(\bk) \tilde{w}_{\mu b}(-\bk) 
\tilde{\cal B}_{\lambda\mu}(k) \times\nonumber\\ 
&&\left(
\frac{\exp[(\zeta_\lambda+\zeta_\mu)\tau]-1}{\zeta_\lambda+\zeta_\mu}
\right)
\label{eq:spd2}
\ea
where we have introduced
\be
\tilde{\cal B}_{\lambda\mu}(k)=\langle \tilde{\bf
v}_\lambda(\bk)|\tilde{\bf B}(k) | \tilde{\bf
v}_\mu(-\bk)\rangle,
\label{eq:cali2}
\ee
and the eigenvalues $\zeta_\lambda(k)$ depend
only
on $|\bk|$ (see App.\ \ref{sec:app2a}). 
The results (\ref{eq:stab}) or (\ref{eq:spd2}) yield the
structure factors, which can be calculated numerically or
analytically.
Expression (\ref{eq:stab}) is most convenient for numerical
computation,
whereas (\ref{eq:spd2}) is more suitable for our theoretical
analysis in
the long wavelength range.
The expressions above contain exponentially growing factors
describing the unstable modes, as in the Cahn-Hilliard (CH) 
theory, discussed in the introduction.
The present theory includes in (\ref{eq:stab})
the rapid microscopic fluctuations,
induced by the Langevin noise $\tilde{\bf C}(k)$, 
and is equivalent to the Cahn-Hilliard-Cook
theory \cite{langer}.
If the Langevin noise is neglected by setting $\tilde{\bf
C}(k)={\bf 0}$ in (\ref{eq:stab}), we obtain the predictions of
the {\em noiseless} CH theory,
\be
\tilde{\bf S}(\bk,\tau)=\exp[\widetilde{\bf M}(\bk) \tau]\cdot
\tilde{\bf S}(\bk,0)\cdot \exp[\widetilde{\bf M}^T(-\bk) \tau].
\label{eq:stabch}
\ee
or in component form
\be
\tilde{S}_{ab}(\bk,\tau)=\sum_{\lambda\mu} \tilde{w}_{\lambda
a}(\bk) \tilde{w}_{\mu b}(-\bk) 
\tilde{\cal S}_{\lambda\mu}(\bk) \exp[(\zeta_\lambda+\zeta_\mu
)\tau],
\label{eq:spd2ch}
\ee
where $\tilde{\cal S}_{\lambda\mu}(\bk)$ is defined by
(\ref{eq:cali2}) with $\tilde{\bf B}(k)$ replaced by $\tilde{\bf
S}(\bk,0)$.

The explicit solutions can be studied using the eigenvalues and
eigenfunctions in different ranges of wave numbers.
For the very long wavelengths in the dissipative range ($k l_0\ll
\gamma_0=\eps/2d$) this will be done in the next section by 
$k$-expansion at fixed $\eps$.
In general, the dissipative range ($k l_0 \ll \gamma_0$) and
the standard range ($k l_0\gg \gamma_0$) are accessible to
analysis by rescaling the wavelengths as $k=\eps^2 \tilde{k}$ or
$k=\sqrt{\eps}\tilde{k}$ respectively, and taking the
small-$\eps$ limit subsequently \cite{mcnamara}.

A theoretical analysis of the formal results, derived in this section,
will be presented in
Sec.\ \ref{sec:devo}.
Before concluding this section we present the numerical
results of (\ref{eq:stab}) and (\ref{eq:stabch}) respectively
with and without Langevin noise, where the relevant equations
have been solved with {\sc Mathematica}.
The values of $S_{ab}(\bk,t)$ `with noise' are plotted as {\em
solid} lines in Figs.\ \ref{fig:snn}, \ref{fig:stl}, and
\ref{fig:sab} for different components
$(ab)$; the values `without noise' are shown as {\em dashed}
lines in Figs.\ \ref{fig:snn}(a) and \ref{fig:stl}. 

The qualitative features of both theories are about the same,
except that the large-$k$ limit of the results of
(\ref{eq:stabch}) `without noise' at fixed $t$ does not
approach the plateau values (\ref{eq:init}), but vanishes.
In the long wavelength and large time limit the results of both
theories approach each other, but at finite $k$ there are
substantial differences.
It should of course be kept in mind that the Cahn-Hilliard
theory without noise was only designed to explain the structure
and instabilities on the largest wavelengths.

It is also of interest to observe that the locations of the
maxima of $S_{nn}$ in Fig.\ \ref{fig:snn} and those of the minima (`dip') 
of $S_\parallel$ in
Fig.\ \ref{fig:stl} approximately coincide.
It indicates that the density instability is closely connected 
to the dynamics controlling $S_{\parallel}$.
This turns out to be the heat mode, as we will show in the next
section.
The negative value of $\tilde{S}_{nT}(k,\tau)$ in Fig.\
\ref{fig:sab} shows that density and temperature fluctuations
are anticorrelated at all wavelengths.

An illustration of the comparison with the MD simulations on systems
of inelastic hard disks of Refs.\ 
\cite{noije-prl,noije-prerc,orza+brito+ernst} is shown in
Fig.\ \ref{fig:snn}(b) for $S_{nn}$ and in Fig.\ \ref{fig:stl}(b)
for $S_\perp\equiv S_{\perp\perp}$ and $S_\parallel\equiv S_{ll}$.
The agreement between theory and simulations is in general very
good, even down to rather large inelasticities ($\alpha \gtrsim
0.6$).
By comparing the simulation results for $S_\perp$ and
$S_\parallel$ in Fig.\ \ref{fig:stl}(b) with the numerical
results of our theory with noise (solid lines) and without
(dashed lines), we have observed that the agreement in the former case
extends over the full range of $k$ values, whereas in the latter
case it is restricted to the small-$k$ range.
Similar conclusions hold when comparing the simulation results
for $S_{nn}(k,t)$ in Fig.\ \ref{fig:snn}(b) with the numerical
results in Fig.\ \ref{fig:snn}(a) with and without Langevin
noise included. 
A more comprehensive comparison with MD simulations for different
densities and different inelasticities will be given in Ref.\
\cite{orza+brito+ernst}, where also the range of validity of the
present theory will be explored.

\section{Development of structure and instabilities}
\label{sec:devo}
An elastic fluid in thermal equilibrium does not show any
structure on {\em hydrodynamic} length scales ($k \lesssim {\rm
min} \{1/l_0,1/\sigma\})$.
This means that the hydrodynamic structure factors $S_{ab}(k)$
are totally flat, independent of $k$, as can be seen in
(\ref{eq:init}). 
The corresponding hydrodynamic correlation functions are
short ranged, $G_{ab}(\br,t)\sim \delta(\br)$,
on these length scales.
Development of structure on length scales above the microscopic
scales $\{l_0,\sigma\}$ will manifest itself in the appearance of
one or more maxima or peaks in the structure factors.
A {\em linear instability} will manifest itself in a structure
factor that grows exponentially in time.
With these concepts in mind, we analyze the structure factors in
(\ref{eq:stab}) for the IHS fluid,
as we want to determine which physical excitations are
responsible for the features observed in the MD simulations and
in the numerical solutions.

\subsection{Structures in the velocity field}
\begin{center}
\begin{figure}[h]
\centerline{\hspace{-8.5cm}\psfig{figure=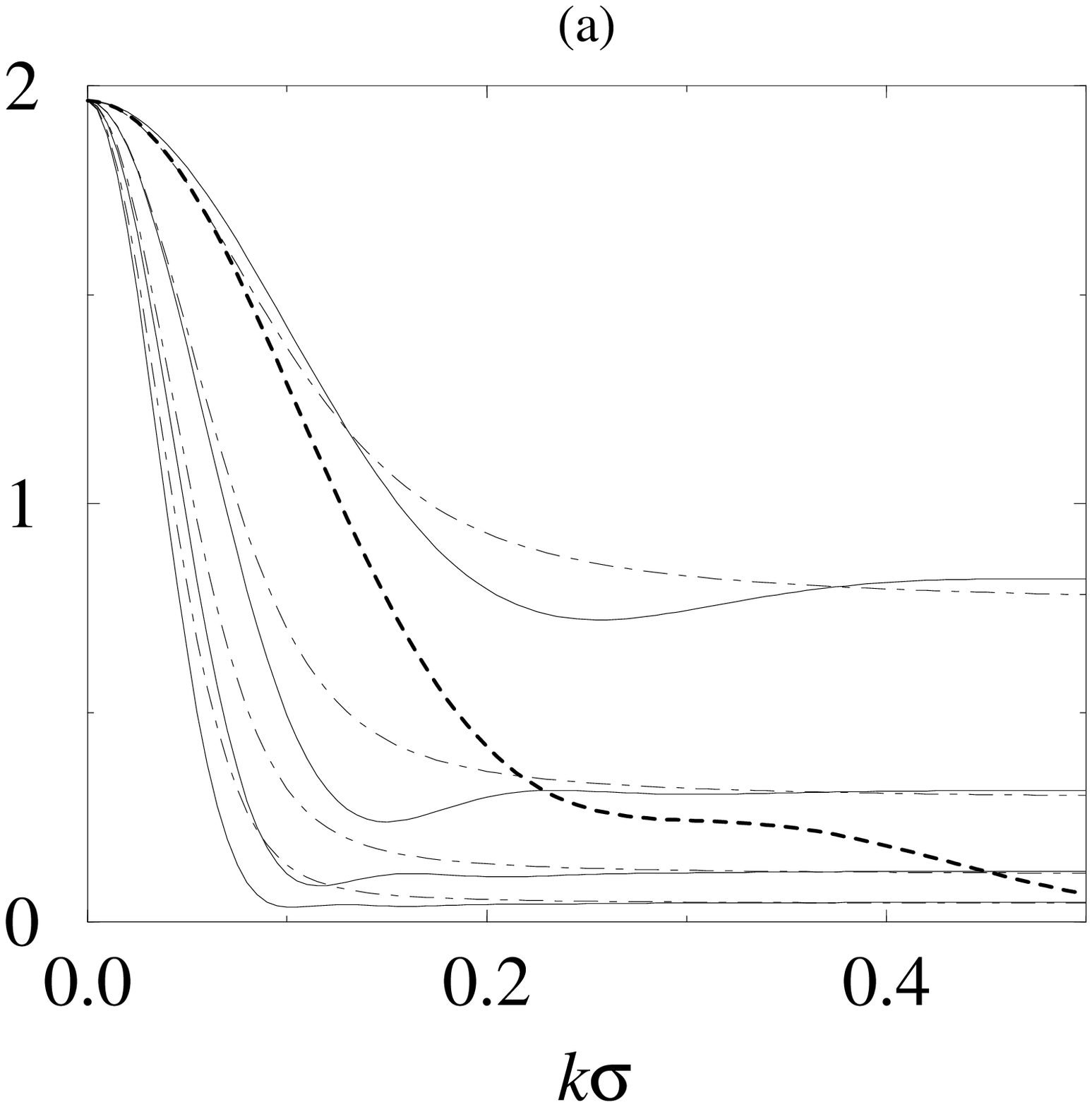,height=7.5cm}}
\vspace{-7.5cm}
\centerline{\hspace{8cm}\psfig{figure=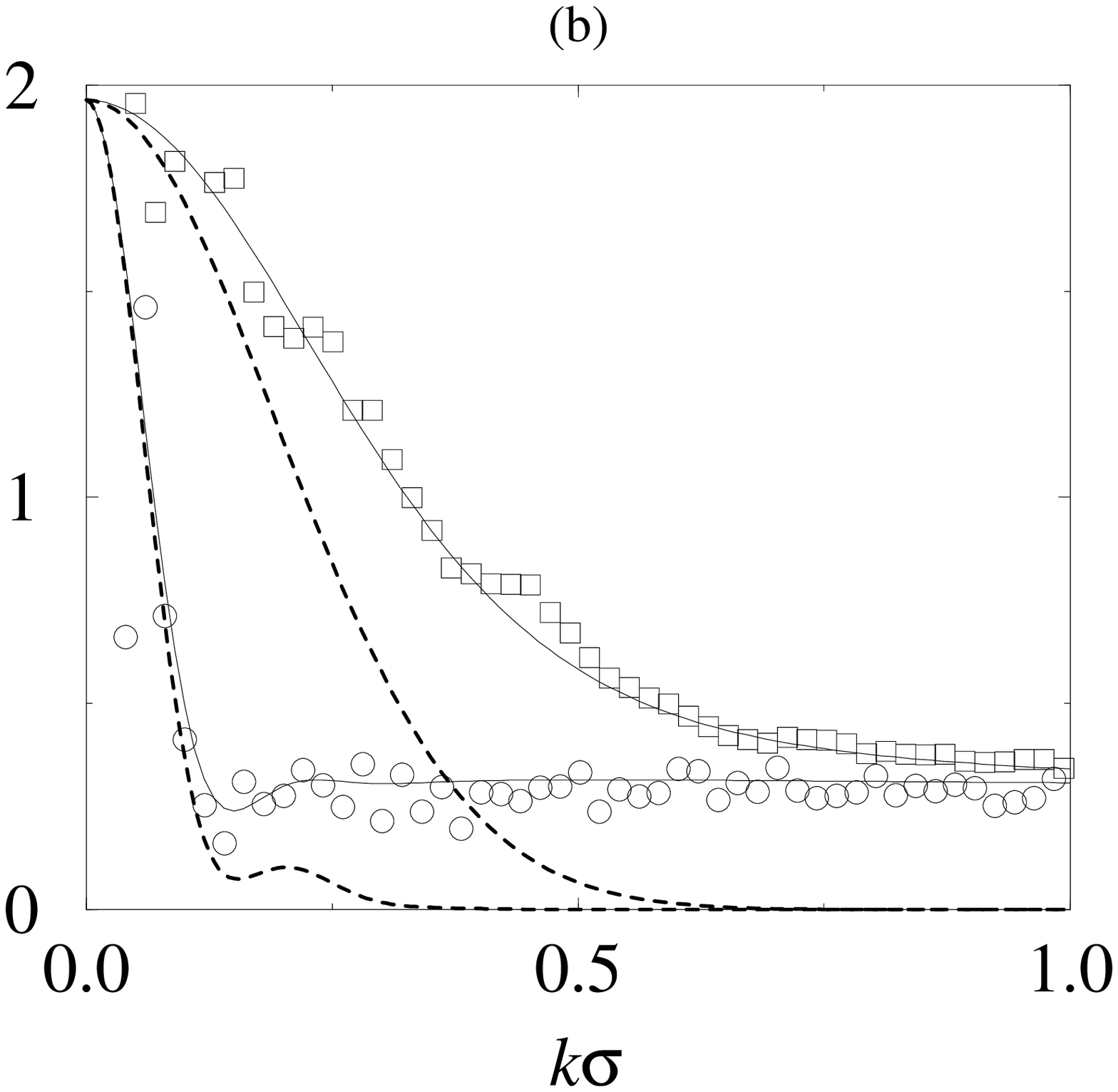,height=7.5cm}}
\vspace{1cm}
\caption{\label{fig:stl}}
\end{figure}
\end{center}

We first consider the simplest case of the 
the transverse structure factor $S_\perp(k,t)=v_0^2(t)
\tilde{S}_\perp(k,\tau)$ with
$(ab)=(\perp\perp)$.
It describes the transverse velocity or vorticity fluctuations
$\tilde{u}_\perp(\bk,\tau)$, which are decoupled in
(\ref{eq:change}) from the remaining Fourier modes, and satisfy
a one-component Langevin equation, where the matrix $\widetilde{\bf  
M}(\bk)$ in (\ref{eq:match}) reduces to a single number 
$\zeta_\perp(k) =\gamma_0(1-\xi_\perp^2 k^2)$.
The structure factor is readily found from (\ref{eq:spd2}) and
yields
\ba
S_\perp^+(k,t)&=&v_0^2(t) \int_0^\tau {\rm d}\tau^\prime
\tilde{B}_{\perp\perp}(k) \exp[2
\zeta_\perp(k)\tau^\prime]\nonumber\\
&=&\frac{T(t)}{m n} \left( \frac{\exp[2\gamma_0(1-\xi_\perp^2
k^2)\tau]-1}{1-\xi_\perp^2 k^2}\right),
\label{eq:sperp2}
\ea
where $\textstyle{\frac{1}{2}}m
v_0^2(t)=T(t)=T_0\exp(-2\gamma_0\tau)$.
Combination of (\ref{eq:dels2}) and (\ref{eq:sperp2}) yields the
complete structure factor, $S_\perp(k,t)=v_0^2(t)[\tilde{S}_\perp(k,0)+\tilde{S}^+_\perp(k,\tau)]$, which is plotted as the upper
{\em solid} line in Fig.\ \ref{fig:stl}(b).
It does not grow, but slowly decays as $S_\perp(k,t)\simeq
(T_0/mn) \exp(-2\gamma_0\xi_\perp^2k^2 \tau)$ at the {\em
largest} wavelengths.
It simply represents vorticity diffusion on the `internal' time
scale $\tau$, with a diffusivity $\gamma_0
\xi_\perp^2=\nu/\omega$, where the typical length scales of
vortices grow \cite{boston} like $L_v(t)\sim 2\pi
\xi_\perp\sqrt{2\gamma_0\tau}\sim 2\pi\sqrt{\nu \tau/\omega}$,
which is independent of the degree of inelasticity.

Consider the {\em noiseless} CH theory in (\ref{eq:stabch}),
which yields directly $S_\perp(k,t)=(T_0/mn)\exp(-2
\gamma_0\xi_\perp^2 k^2 \tau)$ [upper {\em dashed} line in
Fig.\ \ref{fig:stl}(b)].
It agrees with the full theory (\ref{eq:sperp2})
[upper {\em solid} line in Fig.\ \ref{fig:stl}(b)]
only in the small-$k$ limit,
where the denominator $(1-\xi_\perp^2 k^2)$ in
(\ref{eq:sperp2}) can be replaced by 1.
At finite $k$, the differences between both theories are
relatively large.
Moreover, the upper dashed line in Fig.\ \ref{fig:stl}(b) 
convincingly
shows
that the noiseless CH theory does not agree with 
the simulation
results.

Next, we consider the 
longitudinal
structure factor $S_\parallel(k,t)$ with $(ab)=(ll)$.
As subsequent analysis will show, it is determined --- in an
almost quantitative manner --- by a single mode, the heat mode.
The fastest growing term in (\ref{eq:spd2}) is $(\lambda\mu)=(HH)$,
as can be seen from Fig.\ \ref{fig:disp}. 
This leads in the structure factor $S_\parallel(k,t)=v_0^2(t)
\tilde{S}_\parallel(k,\tau)$ to a slowly decaying contribution.
All remaining contributions decay at least as fast as
$v_0(t)\sim \exp(-\gamma_0\tau)$.
The slowest decay occurs at very long wavelengths, i.e. in the
{\em dissipative} range ($k l_0 \ll \gamma_0$).
The relevant coefficient $\tilde{\cal B}_{HH}(k)$ follows from
(\ref{eq:cali2}), (\ref{eq:b2}) and (\ref{eq:eigenve2}) in
App.\ \ref{sec:app2a}, and is given by
\ba
\tilde{\cal B}_{HH}(k)&=&\sum_{ab} \tilde{v}_{Ha}(\bk)\tilde{v}_{Hb}(-\bk)
\tilde{B}_{ab}(k)\nonumber\\
&=&2\gamma_0 \tilde{S}_{ll}(k,0)=\frac{\gamma_0}{n}.
\label{eq:bhh}
\ea
The required component of the right eigenvector, given in
(\ref{eq:eigenve2}),
is $\tilde{w}_{Hl}(\bk) \simeq 1$.
Inserting these data in (\ref{eq:spd2}) and combining it with
(\ref{eq:dels2}) yields for $k l_0\ll \gamma_0$,
\be
S_\parallel(k,t)=\frac{T(t)}{m n} \left(1+
\frac{\exp[2\gamma_0(1-\xi_\parallel^2
k^2)\tau]-1}{1-\xi_\parallel^2 k^2}\right),
\label{eq:spar2}
\ee
where the dispersion relation
$\zeta_H(k)=\gamma_0(1-\xi_\parallel^2 k^2)$ has been used.
This {\em long} wavelength approximation for $S_\parallel$
has the same form as the exact result (\ref{eq:sperp2}) with
$\xi_\perp$ replaced by $\xi_\parallel$.
In Fig.\ \ref{fig:stl}(a) we compare the result (\ref{eq:spar2})
(dot-dashed lines)
with the numerical solution (solid lines), 
presented in Sec. \ref{sec:meso}.
The simple analytic small-$k$ approximation (\ref{eq:spar2})
captures the global features at small $k$ and the plateaux at
larger $k$ values quantitatively, but is missing the little dip
at intermediate $k$ values.

The behavior of the longitudinal structure factor on the {\em
largest} length scales follows again from
Eqs. (\ref{eq:spar2}) as $S_\parallel(k,t)\simeq (T_0/mn)\exp(-2
\gamma_0 \xi_\parallel^2 k^2 \tau)$.
This implies that the heat mode on the largest length scales is
a purely diffusive mode with a diffusivity
$\gamma_0\xi_\parallel^2$, which is much larger than the
diffusivity $\gamma_0 \xi_\perp^2$ for the vorticity 
(see Fig.\ \ref{fig:ratio}), and the associated length scale
grows like $L_\parallel(t)\sim 2\pi\xi_\parallel\sqrt{2\gamma_0\tau}$.
Inspection of the eigenmodes $\{\tilde{\bf w}_H(\bk),\tilde{\bf
v}_H(\bk)\}$ in (\ref{eq:eigenve2}) of App.\ \ref{sec:app2a} for
$k\rightarrow 0$ shows that this diffusive mode is a pure
longitudinal velocity field $\tilde{u}_l(\bk,\tau)$.
Its diffusivity $\gamma_0\xi_\parallel^2$,  given in (\ref{eq:xip})
of App.\ \ref{sec:app2a}, depends for small inelasticities ($\gamma_0\rightarrow 0$) mainly on thermodynamic variables, like
compressibility and pressure, and only slightly on transport
coefficients.

The physical implications of Fig.\ \ref{fig:stl} are quite
interesting.
It shows the phenomenon of {\em noise reduction} \cite{brito+ernst}
at small wavelengths.
With increasing time the noise strengths $S_\parallel(k,t)$ and
$S_\perp(k,t)$ of the fluctuations in the flow field decrease at
larger $k$ values and remain bounded for all $k$ and $t$ by their
initial equipartition value $T_0/m n$, which is independent of $k$.
This can be rephrased by stating that the flow field exhibits only
a `relative' instability.

The noise reduction is a direct consequence of the microscopic
inelastic collision rule, which forces the particles to move more
and more parallel in successive collisions.
It is this `physical coarse graining' process that selectively
suppresses the shorter wavelength fluctuations in the flow field in
an ever-increasing range of wavelengths.
Consistent with this picture is also the selective suppression of
the divergence of the flow field $u_\parallel(\bk,t)$, which
decays at a much faster rate $\gamma_0 \xi_\parallel^2 k^2$ than 
its rotational part $u_\perp(\bk,t)$ that decays 
with a rate $\gamma_0
\xi_\perp^2
k^2$.

So, noise reduction is the pattern selection mechanism, responsible
for the growing vortex structures observed in Fig.\
\ref{fig:patterns}.
An interesting analog from structural geology of
the peak formed at $\bk={\bf 0}$ in reciprocal space, is the formation of
Ayers Rock (Mount Uluru) in the center of Australia, as a result of
peneplanation by selective erosion \cite{geo}.

\subsection{Unstable density structures}
So far, we have seen that the velocity structure factors,
$S_\perp(k,t)$ and $S_\parallel(k,t)$,
develop structure, but remain bounded by their initial values.
Next we will focus on the unstable structure factor
$S_{nn}(k,t)$,
which describes the density clustering in undriven IHS fluids.
In the comparable case of spinodal decomposition the phase
separation is driven by a single unstable mode, the composition
fluctuations, described by the macroscopic equation
$\partial_t \delta n(\bk,\tau)=z_D(k) \delta
n(\bk,t)$, where the growth rate has the typical form
$z_D(k)=\Delta k^2(1-\textstyle{\frac{1}{2}} \xi_D^2 k^2)$.
The corresponding structure factor (\ref{eq:stabch}) in the
noiseless Cahn-Hilliard theory has the form $S_{nn}(k,t)\sim
\exp[2 z_D(k) t]$, and exhibits a maximum growth rate at $k_{\rm
max}=1/\xi_D$, where $z_D(k)$ has a maximum.
This time independent length scale fails to describe the growing
length scales of the patterns observed in spinodal decomposition
\cite{langer}.

However, the present theory does explain the growing correlation
length $L_{\rm cl}(t)$ for the clusters in undriven IHS
fluids in the early stages of cluster formation, as the numerical
results for $S_{nn}(k,t)$ in Fig.\ \ref{fig:snn} demonstrate.
Its maximum at $k_{\rm max}(t)$ shifts to smaller $k$ values.

\begin{center}
\begin{figure}[h]
\centerline{\hspace{-8.5cm}\psfig{figure=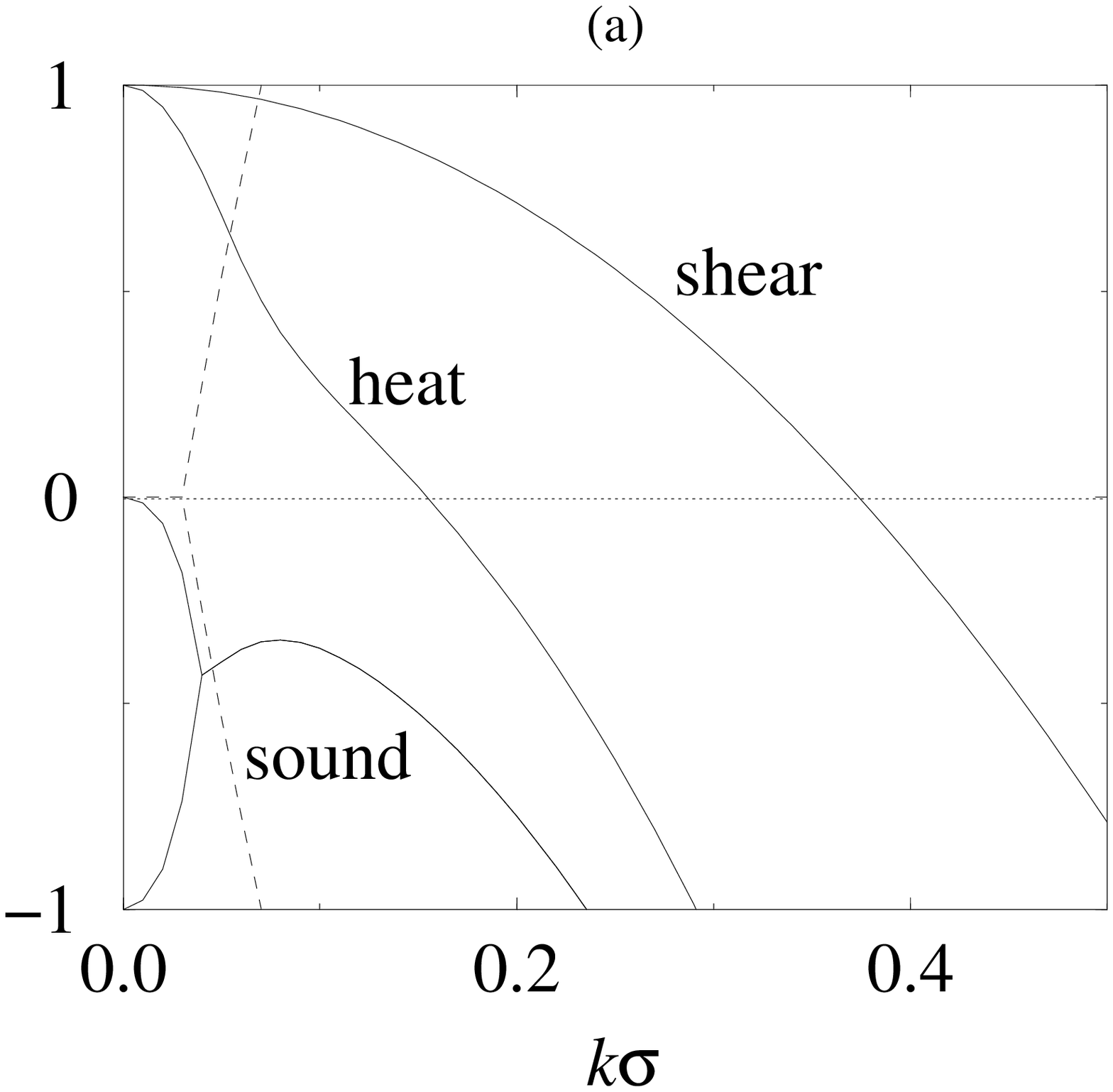,height=7.5cm}}
\vspace{-7.6cm}
\centerline{\hspace{8cm}\psfig{figure=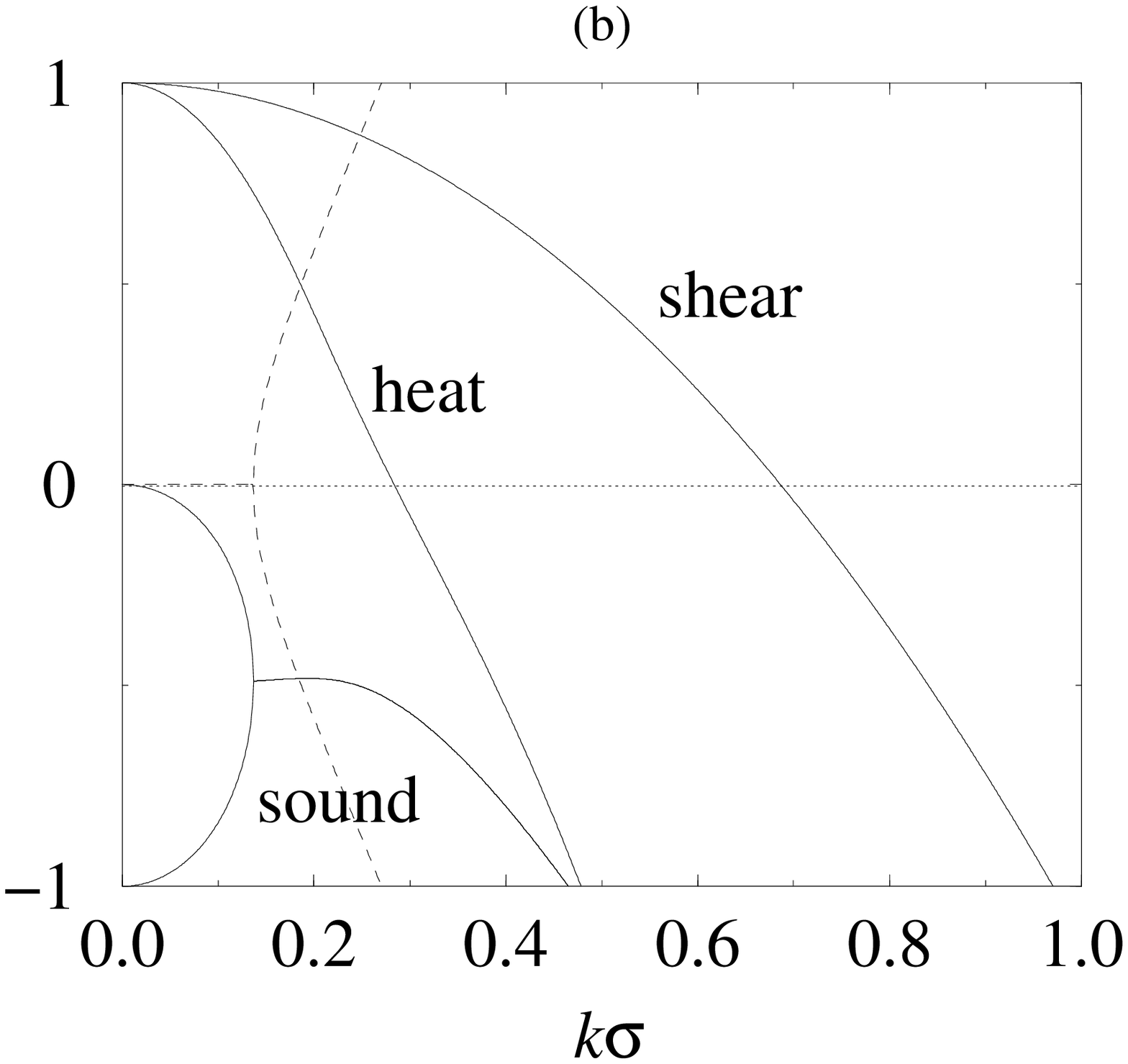,height=7.5cm}}
\vspace{1cm}
\caption{\label{fig:disp}}
\end{figure}
\end{center}

We first consider a naive version of the {\em noiseless}
Cahn-Hilliard theory, proposed by Deltour and Barrat \cite{deltour}.
These authors assume that the structure factor $S_{nn}$ can be
described by the unstable density field, i.e.\ 
$S_{nn}(k,t)\simeq S_{nn}(k,0) \exp[2 \zeta_H(k)\tau]$, with the
growth rate $\zeta_H(k)$ of the unstable heat mode.
As $\zeta_H(k)$ decreases monotonically with $k$, as shown in
Fig.\ \ref{fig:disp}, this structure factor shows the fastest
growth at the {\em smallest} wave number $k_{\rm min}=2\pi/L$,
allowed in a box of length $L$, and does not explain the
dynamics of cluster growth.

Next we consider the full theory of Sec.\ \ref{sec:meso} with
Langevin noise included.
Then the structure factor $S^+_{nn}(k,t)=n^2 
\tilde{S}^+_{nn}(k,\tau)$ is given by (\ref{eq:spd2}) as a sum
over all pairs $(\lambda\mu)$ of hydrodynamic modes, which
exclude shear modes.
Each term in this sum contains a factor
$\exp[(\zeta_\lambda+\zeta_\mu)\tau]$, which leads to
exponential growth when $\zeta_\lambda+\zeta_\mu>0$.
Inspection of the growth rates in Fig.\ \ref{fig:disp} shows
that this happens for $(\lambda\mu)=(HH)$ for all $k<k_H^\ast$
[defined in App.\ \ref{sec:app2a} below (\ref{eq:eigenve2})],
and for $(\lambda\mu)=(H+)$ for $k$ below a certain threshold.
The label $\lambda=+$ refers to the mode with vanishing rate
constant as $k\rightarrow 0$ (see App.\ \ref{sec:app2a}).

To obtain a {\em qualitative} understanding of the clustering
instability, we present an analysis based on the most unstable
pair of modes, $(\lambda\mu)=(HH)$.
In this approximation we obtain from (\ref{eq:spd2}),
\ba
S^+_{nn}(k,\tau)&=&n^2
\tilde{w}_{H
n}(\bk) \tilde{w}_{Hn}(-\bk)
\tilde{\cal B}_{HH}(k)\times\nonumber\\ 
&&\left(
\frac{\exp[2\zeta_H(k)\tau]-1}{2\zeta_H(k)}.
\right)
\label{eq:spdnn2}
\ea
We analyze this expression in the dissipative range, $k
l_0\ll \gamma_0$, where $\tilde{\cal B}_{HH}(k)$ is given
in (\ref{eq:bhh}), and where $\zeta_H(k)\simeq
\gamma_0(1-\xi_\parallel^2
k^2)$.
The component $\tilde{w}_{Hn}(\bk)$ in (\ref{eq:eigenve2}) of
App.\ \ref{sec:app2a} vanishes for small $k$.
Explicit calculation to the next order in $k$,
as given in Ref.\ \cite{brito+ernst},
shows that $\tilde{w}_{Hn}(\bk) = -i k l_0/\gamma_0$.
These results combined with (\ref{eq:dels2}) yield 
for the structure factor in the dissipative $k$
range,
\be
S_{nn}(k,t)= n^2 T \chi_T + \frac{n k^2
l_0^2}{2\gamma_0^2}
\left(
\frac{\exp[2\gamma_0(1-\xi_\parallel^2
k^2)\tau]-1}{1-\xi_\parallel^2 k^2}\right).
\label{eq:snna}
\ee
This simple analytic approximation agrees within the small-$k$
range for which it is derived (say up to $k\sigma \lesssim 0.1$
in Fig.\ \ref{fig:snn}), with the numerical solutions of the
theory either with or without Langevin noise.
Moreover, it demonstrates that the instability is driven through a 
coupling to 
the unstable `heat' mode, which is, in the small-$k$ range, a
longitudinal velocity mode.
The coupling of the density structure factor to the unstable heat
mode is 
rather weak,
${\cal O}(k^2)$, which explains why structure in the flow field
appears long before density clusters appear.

The wave number $k_{\rm max}(t)$ of the maximum growth of
$S_{nn}$ in (\ref{eq:snna}) determines the typical length scale
of the density clusters.
For $2\gamma_0\tau \gg 1$, it can be determined analytically as
$L_{\rm cl}(t)\sim 2\pi/k_{\rm max}(t)=2\pi\xi_\parallel\sqrt{2\gamma_0\tau}$,
which is the same length scale, $L_\parallel(t)$, as appeared in
$S_\parallel$.
The good agreement between theory and MD simulations, shown in
Fig.\ \ref{fig:snn}, confirm that the {\em initial} growth of
density inhomogeneities is indeed controlled by the {\em
longitudinal} flow field with a length scale $L_{\rm cl}(t)\sim
\sqrt{\tau/\eps}$ at small inelasticity $\eps=2d\gamma_0$,
and {\em not} by the transverse flow field with a length scale
$L_v(t)\sim 2\pi\xi_\perp \sqrt{2\gamma_0\tau} \sim \sqrt{\tau}$,
independent of $\eps$.
The pattern selection mechanism for the vortex structures is very
different from the mechanism that leads to the formation of density
clusters.
This is the more common linear instability in density or
composition fluctuations, which also occurs in spinodal
decomposition \cite{langer}.

\section{Spatial correlation functions}
\label{sec:spatial2}
Once the structure factors have been obtained, the correlation 
functions
can be calculated by Fourier inversion.
When $a,b$ refer to $n$ and $T$ the components of
$S_{ab}(\bk,t)$
and $G_{ab}(\br,t)$ are scalar isotropic functions only
depending
on $|\bk|$ and $|\br|$ respectively.
When $(a,b)=(\alpha,\beta)$ refer to Cartesian components
$u_\alpha$
of the flow field, then $S_{\alpha\beta}(\bk,t)$ is a second
rank isotropic tensor field, which can be separated into two
independent isotropic scalar functions:
\be
S_{\alpha\beta}({\bf k},t)=\hat{k}_\alpha\hat{k}_\beta
S_\parallel(k,t)+(\delta_{\alpha\beta}-\hat{k}_\alpha\hat{k}_\beta)
S_\perp(k,t),
\label{eq:sdec}
\ee
where $S_\parallel(k,t)$ and $S_\perp(k,t)$ are given by
(\ref{eq:sdef}) with $\delta a$ and $\delta b$ equal to $u_l$
and
$u_\perp$ respectively.
A similar separation applies to the velocity correlation
functions,
\ba
G_{\alpha\beta}(\br,t)&=&V^{-1} \sum_{\bk} \exp(i\bk\cdot \br)
S_{\alpha\beta}(\bk,t)\nonumber\\
&=&\hat{r}_\alpha\hat{r}_\beta
G_\parallel(r,t)+(\delta_{\alpha\beta}-\hat{r}_\alpha
\hat{r}_\beta)
G_\perp(r,t).
\label{eq:gdec}
\ea
Firstly we note that the Fourier series in (\ref{eq:gdec}) can
be
replaced by a Fourier integral, provided that the system is
sufficiently large.
Then, for periodic boundary conditions, as used in MD
simulations, $V^{-1}\sum_{\bk}$ can be replaced by
$(2\pi)^{-d}\int{\rm d}\bk$.
Strictly speaking, isotropic symmetry and separation of
second rank tensor functions into two scalar functions only hold
in the thermodynamic limit.

Secondly, the inverse Fourier transform ${\bf G}(\br,t)$ of
${\bf S}(\bk,t)$ only exists as a {\em classical function}, if
${\bf S}^{\infty}(t)\equiv \lim_{k\rightarrow \infty} {\bf S}(\bk,t)$ vanishes.
If ${\bf S}(\bk,t)$ approaches a nonvanishing constant ${\bf S}^{\infty}(t)$, it
yields a {\em distribution} $\delta(\br)$.
So,
\be
{\bf G}(\br,t)={\bf S}^{\infty} \delta(\br) +\int\frac{{\rm d}\bk}{(2\pi)^d}
\exp(i \bk\cdot\br) {\bf S}^+(\bk,t).
\label{eq:gdel}
\ee
Note that the limiting values $\tilde{S}_{ab}^\infty$, when
expressed in terms of the rescaled variables, are given by the
{\em time} and (wave number) independent values $\tilde{S}_{ab}(k,0)$.
This is the reason for using the same notation as in Eq.\ 
(\ref{eq:dels2}).
In the sequel it is convenient to also use the notation,
\be
{\bf G}^+(\br,t)={\bf G}(\br,t)-{\bf S}^{\infty}(t) \delta(\br).
\label{eq:splus}
\ee
The structure factors $\tilde{\bf S}^+(\bk,\tau)$ in
(\ref{eq:intds}), calculated in the theory {\em with} Langevin
noise, as well as $\tilde{\bf S}(\bk,\tau)$ in (\ref{eq:stabch})
{\em without} noise are vanishing for large $k$, and can be
Fourier inverted.
However, the functions $\tilde{\bf S}(\bk,\tau)$ in (\ref{eq:dels2})
contain a part $\tilde{\bf S}(\bk,0)$, which is {\em independent} of
$k$ in the relevant $k$ interval, and which yields after Fourier
inversion a contribution proportional to $\delta(\br)$.
These `large $k$' contributions are in fact the correlation
functions of an elastic hard sphere (EHS) fluid as $k\rightarrow
0$, i.e.\
\ba
G_{nn}(\br,t)&\simeq&n^2 T \chi_T \delta(\br)\nonumber\\
G_{TT}(\br,t)&\simeq&\frac{2T^2(t)}{dn}\delta(\br)\nonumber\\
G_{\alpha\beta}(\br,t)&\simeq&\frac{T(t)}{mn}\delta(\br)
\delta_{\alpha\beta}.
\label{eq:delta}
\ea
According to Sec.\ \ref{sec:hydrostab}, `large $k$' here means
$\sqrt{\gamma_0}/l_0 \ll k \ll {\rm
min}\{2\pi/l_0,2\pi/\sigma\}$.
Here $G_{nn}$ is the coarse grained density-density correlation
function for EHS, in which the Fourier components with $k\sigma
\gtrsim 2\pi$ have been discarded.
In App.\ \ref{sec:app2b} we derive the formulas, necessary for
the analytic and numerical Fourier inversion of
${\bf S}^+(\bk,t)$, as defined in (\ref{eq:splus}).

\subsection{Incompressible limit}
\label{subsec:incomp}
There is an interesting limiting case of the theory, the
incompressible limit \cite{noije-prl}, 
that greatly simplifies and elucidates the
analytic solution of the full set of coupled linearized equations
(\ref{eq:change}) for hydrodynamic fluctuations.
It is well known from fluid dynamics and the theory of turbulence
\cite{batchelor,frisch} that ordinary elastic fluid flows are
quite incompressible, which implies that $\bnabla\cdot{\bf u}=0$
and that the longitudinal mode $u_l(\bk,t)\simeq 0$.
Then, the nonlinear Eq.\ (\ref{eq:hydrofree}) for the transverse
flow field or, equivalently, for the vorticity, practically
decouples from the remaining hydrodynamic equations.
In the comoving reference frame there is only a {\em nonlinear}
coupling of the temperature fluctuations to the transverse flow
field through the nonlinear viscous heating, $\eta|\bnabla {\bf
u}|^2$.
We therefore expect that the IHS fluid in the nearly elastic case
can be considered as incompressible, at least to lowest
approximation.

What are the consequences of this assumption for the Fourier
modes?
The structure factor $S_{nn}(k,t)$ of density fluctuations does not
evolve in time.
The temperature fluctuation $\delta
T(\bk,t)$ in (\ref{eq:change}) simply decays as a kinetic mode
and the average temperature stays spatially homogeneous.
Clearly, the assumption is too drastic a simplification to
describe the clustering instability.
However, an approximate theory based on vorticity fluctuations
alone is justified to describe the patterns in the flow field,
as discussed in Sec.\ \ref{sec:devo}.

So, we combine the assumption of incompressibility, 
$S_\parallel(k,t)=0$,
with $S_\perp(k,t)$ in (\ref{eq:sperp2}), using
(\ref{eq:sdec}) and (\ref{eq:gdec}).
This enables us, for
thermodynamically large systems, to explicitly calculate the
correlation functions $G_{\alpha\beta}(\br,t)$ of the velocity
field, by inverse Fourier transformation, i.e.\ 
\ba
G_{\alpha\beta}^+(\br,t)&=&\hat{r}_\alpha\hat{r}_\beta
G_\parallel^+(r,t)+(\delta_{\alpha\beta}-\hat{r}_\alpha\hat{r}_\beta) 
G_\perp^+(r,t)\nonumber\\
&&\!\!\!\!\!\!\!\!\!\!\!\!\!\!=\int\frac{{\rm d}\bk}{(2\pi)^d} \exp(i\bk\cdot \br) 
(\delta_{\alpha\beta}-\hat{k}_\alpha\hat{k}_\beta) S^+_\perp(k,t).
\ea
The Fourier transform 
has been calculated in App.\
\ref{sec:app2b}, and yields for the two scalar functions
$G^+_\lambda(r,t)$ with $\lambda=\{\parallel,\perp\}$
\be
G^+_\lambda(r,t)=\frac{T(t)}{mn \xi_\perp^d} 
g_\lambda\left(\frac{r}{\xi_\perp},
2\gamma_0 \tau\right),
\label{eq:glambda}
\ee
where $g_\lambda(x,s)$ is given in (\ref{eq:glamb}) of App.\
\ref{sec:app2b}, and $\gamma_0$ and $\xi_\perp$ are defined below
Eqs.\ (\ref{eq:Gloss}) and (\ref{eq:change}) respectively.

\begin{center}
\begin{figure}[h]
\epsfig{figure=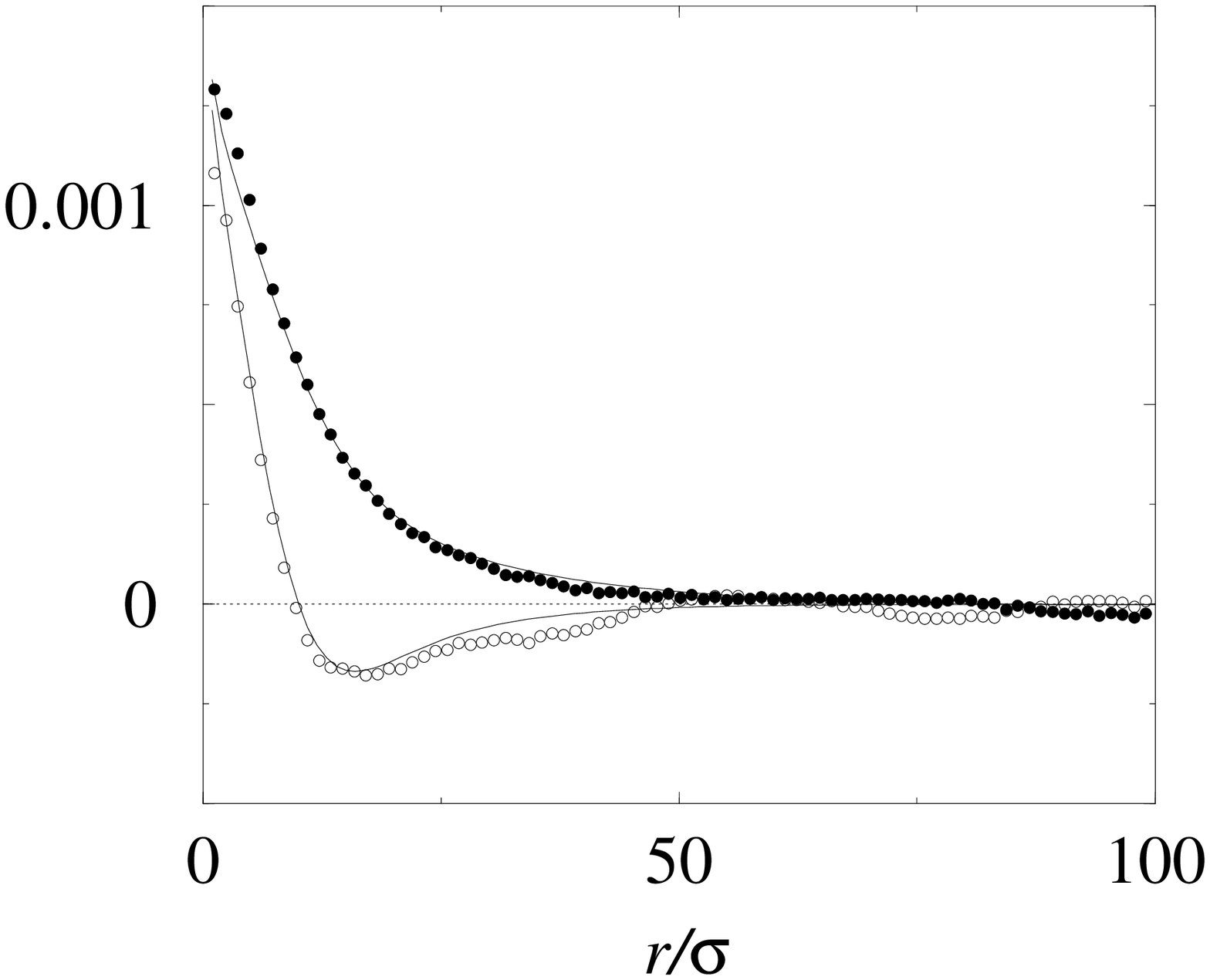,width=10cm}
\caption{\label{fig:gtl.04.09.40}}
\vspace{1cm}
\end{figure}
\end{center}

As an explicit example we show the result for inelastic hard
disks, which is most
relevant for a comparison with existing computer simulations,
in Fig.\ \ref{fig:gtl.04.09.40}.
Their analytic form is,
\ba
g_\parallel(x,s)&=&\frac{1}{2\pi x^2}\int_0^s {\rm d}s^\prime
\exp(s^\prime)
[1-\exp(-x^2/4 s^\prime)]\nonumber\\
g_\perp(x,s)&=& -g_\parallel(x,s)+\int_0^s {\rm d}s^\prime
\exp(s^\prime) \frac{\exp(-x^2/4 s^\prime)}{4\pi s^\prime}.
\label{eq:g2d}
\ea
The function $g_\perp(x,s)$
has a negative minimum,
while $g_\parallel(x,s)$ is positive for all $x,s,d$; 
there are {\em algebraic} tails $g_\parallel(x,s)\sim
-(d-1)g_\perp(x,s)\sim x^{-d}$
with a correction term of ${\cal
O}(\exp(-x^2/4s))$, explicitly given in (\ref{eq:tail}).
Similar algebraic tails occur in nonequilibrium stationary states
in driven diffusive systems \cite{grinstein,schmittmann+zia}.
These functions have structure on hydrodynamic space and time
scales
where both $x=r/\xi_\perp$ and $s=2\gamma_0\tau$ can be
either large or
small with respect to unity.
At small inelasticity ($\gamma_0\rightarrow 0$) the dynamic
correlation length $\xi_\perp$
and mean free path $l_0$ are well separated.

A more systematic comparison between the theoretical predictions 
(\ref{eq:g2d})
and molecular dynamics simulations is made
in Ref.\ \cite{noije-prl,orza+brito+ernst}.
In Fig.\ \ref{fig:gtl.04.09.40}, we show the results from a
single simulation run at packing 
fraction
$\phi=0.4$, and small inelasticity
$\alpha=0.9$.
The parallel part $G_\parallel(r,t)$ exhibits a tail $\sim r^{-d}$
(see Fig.\ 2(a) in Ref.\ \cite{noije-prl})
and shows good agreement, well beyond the crossover time $\tau_c$
(defined in Fig.\ \ref{fig:cooling})
that separates the linear regime from the nonlinear
clustering regime.
The minimum in $G_\perp(r,t)$ at $L_v(t)$ can be identified as
the mean diameter of vortices, shown in Fig.\ \ref{fig:patterns}.
The analytic result for $G_\perp(r,t)$ in (\ref{eq:g2d}) for
large times shows that 
that $L_v(t)\sim 2\pi \xi_\perp\sqrt{2 \gamma_0 \tau}$ is growing
through vorticity diffusion.

Apart from the restrictions to hydrodynamic space and
time scales, there are two essential criteria
limiting the validity of the incompressible theory: 
(i) System sizes $L$ must be
{\em thermodynamically large} ($L\gg 2\pi\xi_\perp$),
so that Fourier sums over
${\bf k}$-space can be replaced by ${\bf k}$-integrals.
(ii) Times must be restricted to the {\em linear}
hydrodynamic regime ($\tau\lesssim\tau_c$), so that the
system 
remains close to the HCS.
It appears that 
our description of the fluctuations in terms of a Langevin
equation
based on incompressibility is confirmed by the
simulations in the linear regime $\tau< \tau_c$
for small inelasticities.

\subsection{Compressible Flows}
\label{subsec:comp}
In this subsection we extend the theory to compressible flows
\cite{noije-prerc}.
The description of the velocity fluctuations
$G_{\alpha\beta}(\br,t)$ in the previous subsection was based on
fluctuating
hydrodynamics for the vorticity fluctuations only, i.e.\ the
absence
of longitudinal fluctuations (incompressibility assumption).
Fig.\ \ref{fig:stl}(b) confirms that this assumption is very 
reasonable
indeed,
as $S^+_\parallel(k,t)=S_\parallel(k,t)-T/mn$  
is vanishingly small down to very small
$k$ values.
However for the smallest wave numbers, the
incompressibility assumption breaks down.
As the analysis of (\ref{eq:sperp2}) and (\ref{eq:spar2}), as
well as the numerical evaluation in Fig.\ \ref{fig:stl} show,
the structure factor
$S_\parallel(k\rightarrow
0,t)=S_\perp(k\rightarrow
0,t)$.
This implies
for large distances 
$G_{\alpha\beta}(\br,t)\sim S_\perp(k\rightarrow 0,t)
\delta_{\alpha\beta} \delta(\br)$, and thus the absence of
algebraic long range correlations on the largest scales ($r\gg
2\pi \xi_\parallel$).
Therefore, we can already conclude that the asymptotic behavior
of
$G_\perp(r,t)$ and $G_\parallel(r,t)$ cannot be $r^{-d}$.
Instead
the $r^{-d}$ tail, obtained in the previous subsection, describes
intermediate
behavior which is exponentially cut off at
a distance determined by the width of $S^+_\parallel(k,t)$.
This width can be estimated from the eigenvalues of the
hydrodynamic matrix, more precisely from the dispersion relation
of
the heat mode,
which is a pure longitudinal velocity $\tilde{u}_l$ for $k\rightarrow 0$.
To second order in $k$ its dispersion relation
is given by $\zeta_H(k)=
\gamma_0(1-k^2 \xi_\parallel^2)$, where $\xi_\parallel$ is given
in Eq.\ (\ref{eq:xip}) of App.\ \ref{sec:app2a}.
Note that for small inelasticities $\xi_\parallel$ and
$\xi_\perp$ are well separated, as $\xi_\parallel\sim 1/\eps$,
whereas $\xi_\perp\sim \xi_l\sim \xi_T \sim 1/\sqrt{\eps}$.

\begin{center}
\begin{figure}[h]
\centerline{\hspace{-9cm}\psfig{figure=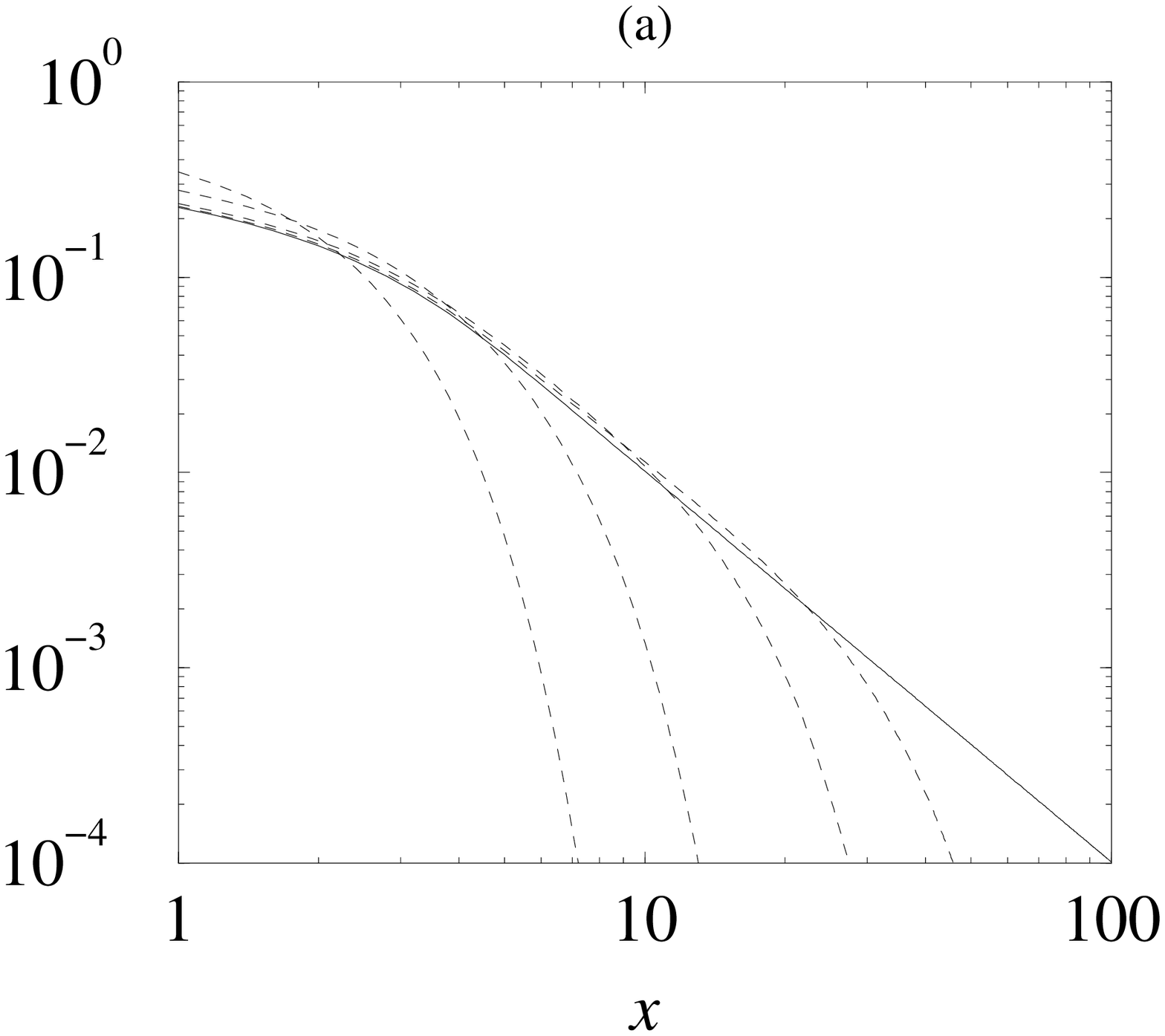,height=7.5cm}}
\vspace*{-7.5cm}
\centerline{\hspace{8cm}\psfig{figure=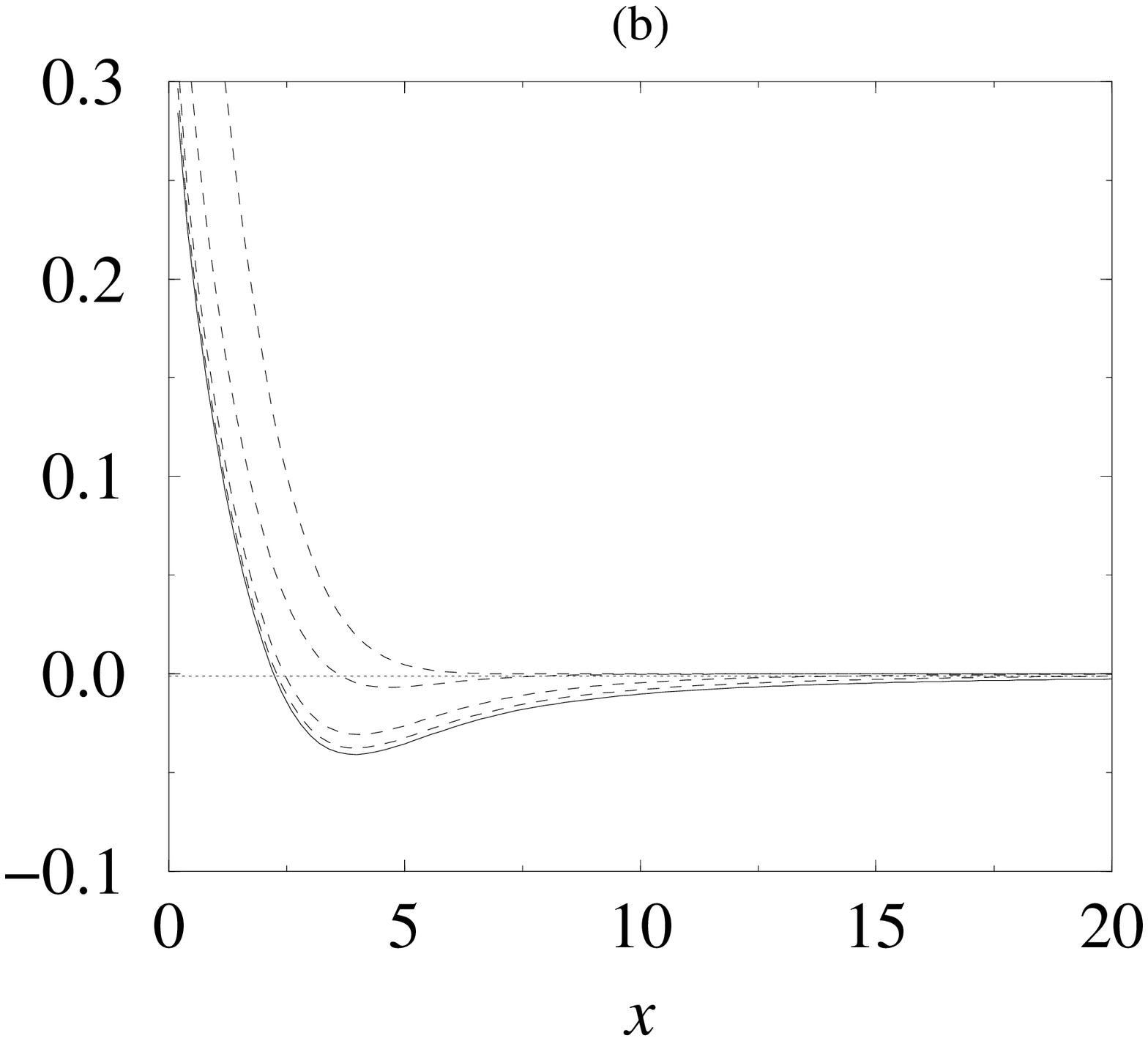,height=7.5cm}}
\vspace{1cm}
\caption{\label{fig:comp}}
\end{figure}
\end{center}

Using approximation (\ref{eq:spar2}) for $S_\parallel(k,t)$,
the structure factor
$S^+_{\alpha\beta}(\bk,t)$ can be written as
\ba
S^+_{\alpha\beta}(\bk,t)&\approx&\frac{T(t)}{m n}\int_0^s
{\rm d} s^\prime \exp(s^\prime)
\left[\hat{k}_\alpha\hat{k}_\beta \exp(-s^\prime
k^2\xi_\parallel^2) \right.
\nonumber\\
&&\left.+(\delta_{\alpha\beta}-\hat{k}_\alpha\hat{k}_\beta)\exp(-
s^\prime k^2 \xi_\perp^2)\right],
\ea
where $s=2\gamma_0 \tau$.
If the system is thermodynamically large ($L\gg 2\pi
\xi_\parallel$),
$G^+_\parallel(r,t)$ and
$G^+_\perp(r,t)$
can be obtained by performing
integrals over $\bk$ space and are
expressed as
integrals over simple functions, as derived in App.\
\ref{sec:app2b}.
Here we only quote the results for $d=2$:
\ba
&&G^+_\lambda(r,t)\approx \frac{T(t)}{mn}\left(\frac{1}{4\pi
\xi_\lambda^2} \int_0^s {\rm d}s^\prime\frac{
\exp(s^\prime -x_\lambda^2/4
s^\prime)}{s^\prime}\right.\nonumber\\
&&\left.+\frac{\sigma_\lambda}{2\pi r^2} \int_0^s {\rm d}
s^\prime e^{s^\prime}\left[\exp\left(-\frac{x_\parallel^2}
{4 s^\prime}\right)-
\exp\left(-\frac{x_\perp^2}{4
s^\prime}\right)\right]\right).
\label{eq:expl}
\ea
for $\lambda=\parallel,\perp$, where $x_\lambda=r/\xi_\lambda$,
$\sigma_\parallel=1$ and $\sigma_\perp=-1$.
We first observe that, in the time regime $\tau\sim 1/\gamma_0$,
$G^+_\lambda(r,t)$ has structure both on the scale $r\sim 2\pi
\xi_\perp$ as well as on $r\sim 2\pi \xi_\parallel$.
Moreover, 
$G_\parallel^+(r,t)$ is {\em positive} both
in the incompressible as well as in the compressible case because
$\xi_\perp < \xi_\parallel$.
In Fig.\ \ref{fig:comp} we show the above approximations for
different values of the ratio $\xi_\parallel/\xi_\perp$,
together with the incompressible limit result of the previous
section, which is 
obtained for $\xi_\parallel\rightarrow
\infty$.
At finite $\eps$, Eq.\ (\ref{eq:expl}) describes exponentially
decaying
functions
at distances $r\gtrsim
2\pi \xi_\parallel$.
Moreover, upon increasing the inelasticity
the minimum in $G_\perp(r,t)$ becomes less
deep and vanishes if $\xi_\parallel=\xi_\perp$.

The predicted spatial velocity correlations
$G_\parallel(r,t)$ and $G_\perp(r,t)$
have been obtained
by performing inverse Bessel
transformations on the numerical results for $S_\parallel(k,t)$
and
$S_\perp(k,t)$.
At small inelasticity ($\alpha \gtrsim 0.9$) the functions
$G_\parallel(r,t)$ and $G_\perp(r,t)$, calculated
from the full set of hydrodynamic equations, differ for
$r\lesssim 2\pi \xi_\parallel$ only slightly from the results for
incompressible flow fields (see discussion in the previous
section).
However, the algebraic tails $\sim r^{-d}$ in $G_\parallel(r,t)$
and $G_\perp(r,t)$ for $r\gtrsim
2\pi \xi_\perp$, as derived in the previous subsection, 
are exponentially cut off for $r\gtrsim
2\pi\xi_\parallel$, as implied by Eq.\ (\ref{eq:expl}).
As the correlation lengths $\xi_\perp\sim 1/\sqrt{\eps}$ and
$\xi_\parallel\sim 1/\eps$ are well separated for small
$\eps$, there is an intermediate range
of $r$ values where the algebraic tail $\sim r^{-d}$ in
$G_\parallel(r,t)$ can be observed.
 
\begin{center}
\begin{figure}[h]
\epsfig{figure=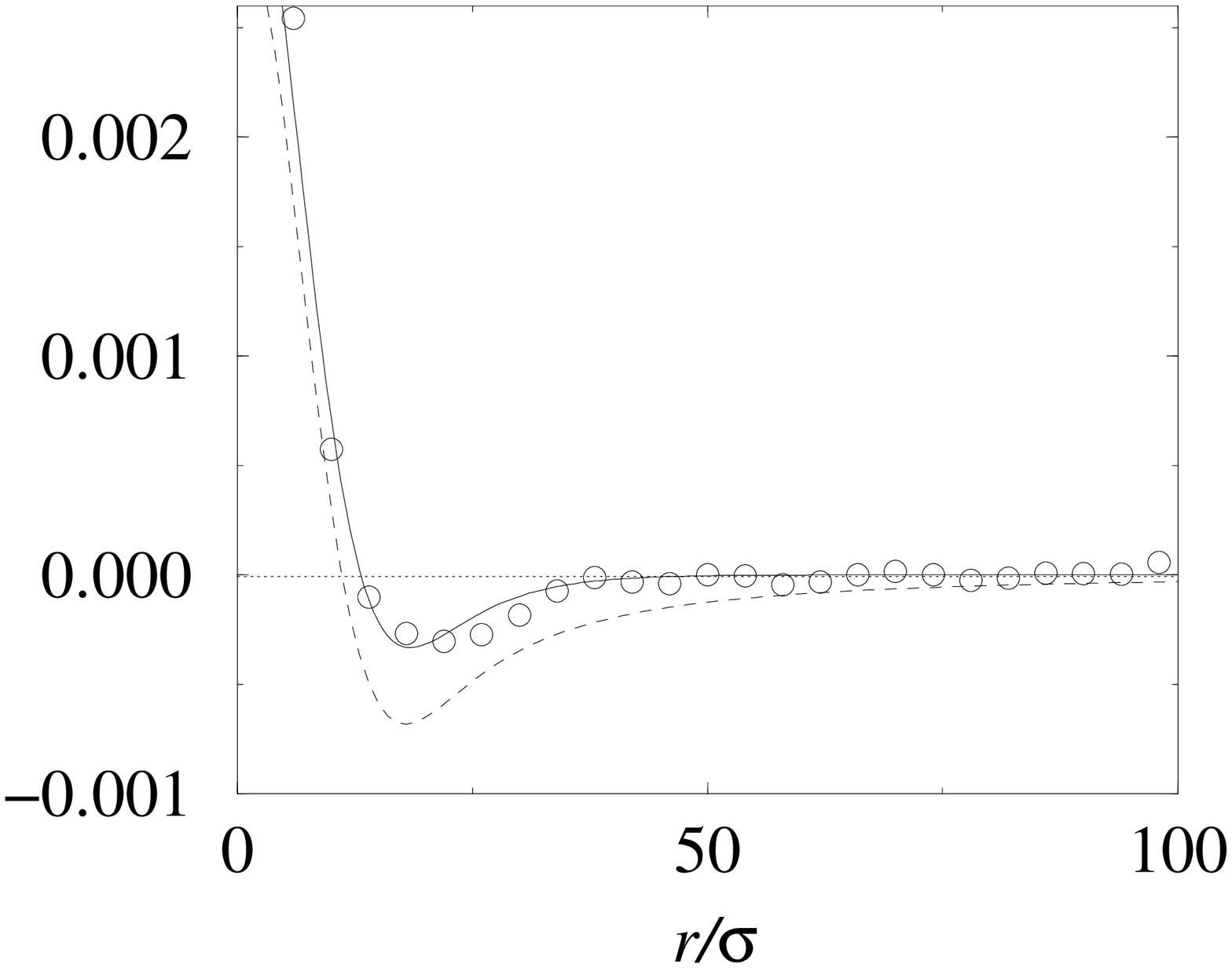,width=10cm,angle=0}
\vspace{1cm}
\caption{\label{fig:gt.04.06.40}}
\end{figure}
\end{center}

At higher inelasticity $\xi_\parallel$ and $\xi_\perp$ are not
well
separated and, as a consequence, there does {\em not} exist a spatial
regime in which the longitudinal fluctuations in the flow field
can
be neglected and the regime of validity of the incompressible
theory
has shrunk to zero.
Fig.\ \ref{fig:gt.04.06.40} compares results 
from incompressible and 
compressible
fluctuating hydrodynamics with simulation data for $G_\perp(r,t)$
at $\phi=0.4$ and $\alpha=0.6$,
and confirms the necessity of including longitudinal velocity
fluctuations to calculate the spatial velocity correlations
at reasonably large inelasticities.

\begin{center}
\begin{figure}[h]
\centerline{\hspace{-8.5cm}\psfig{figure=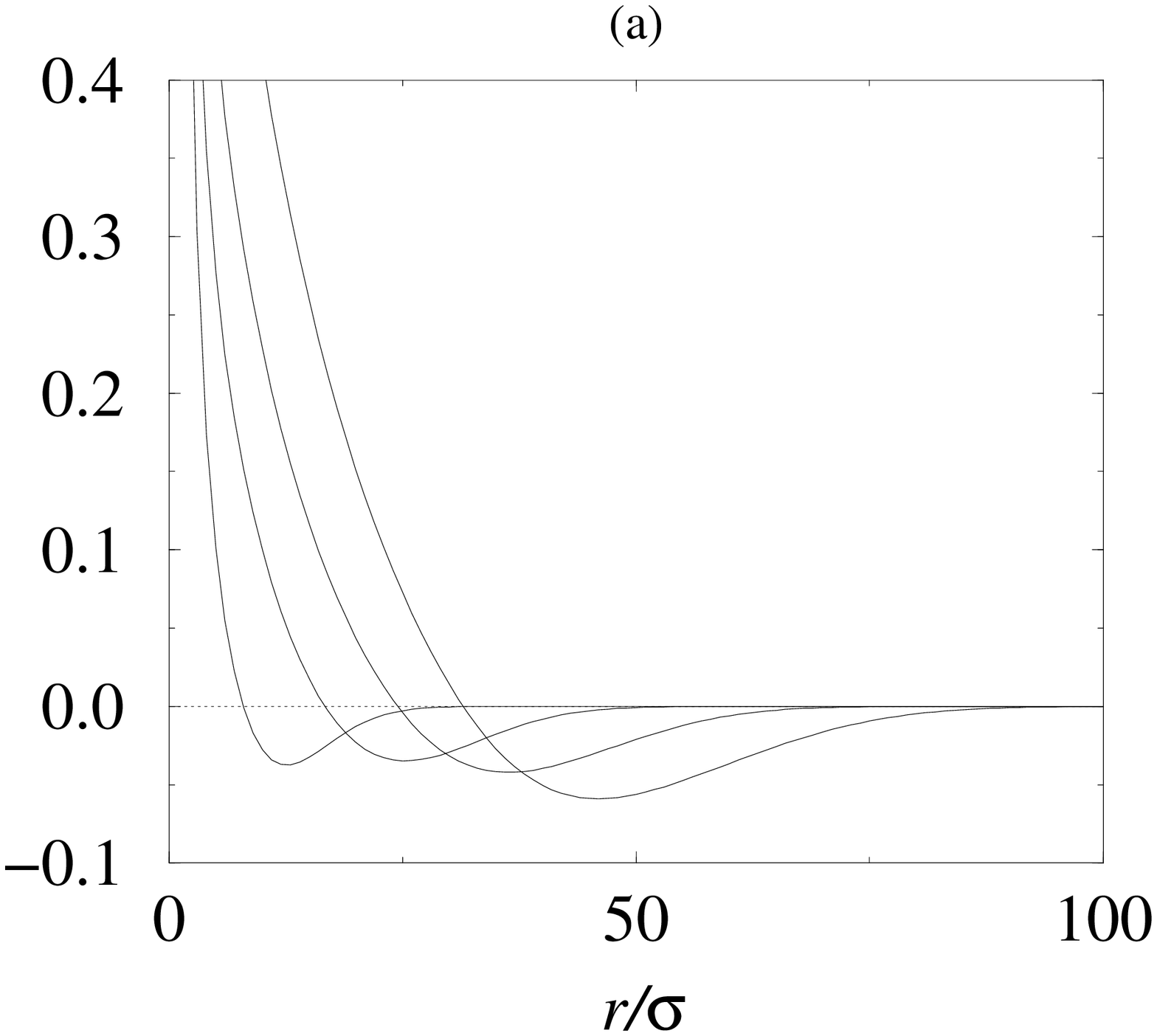,height=7.5cm}}
\vspace{-7.5cm}
\centerline{\hspace{8cm}\psfig{figure=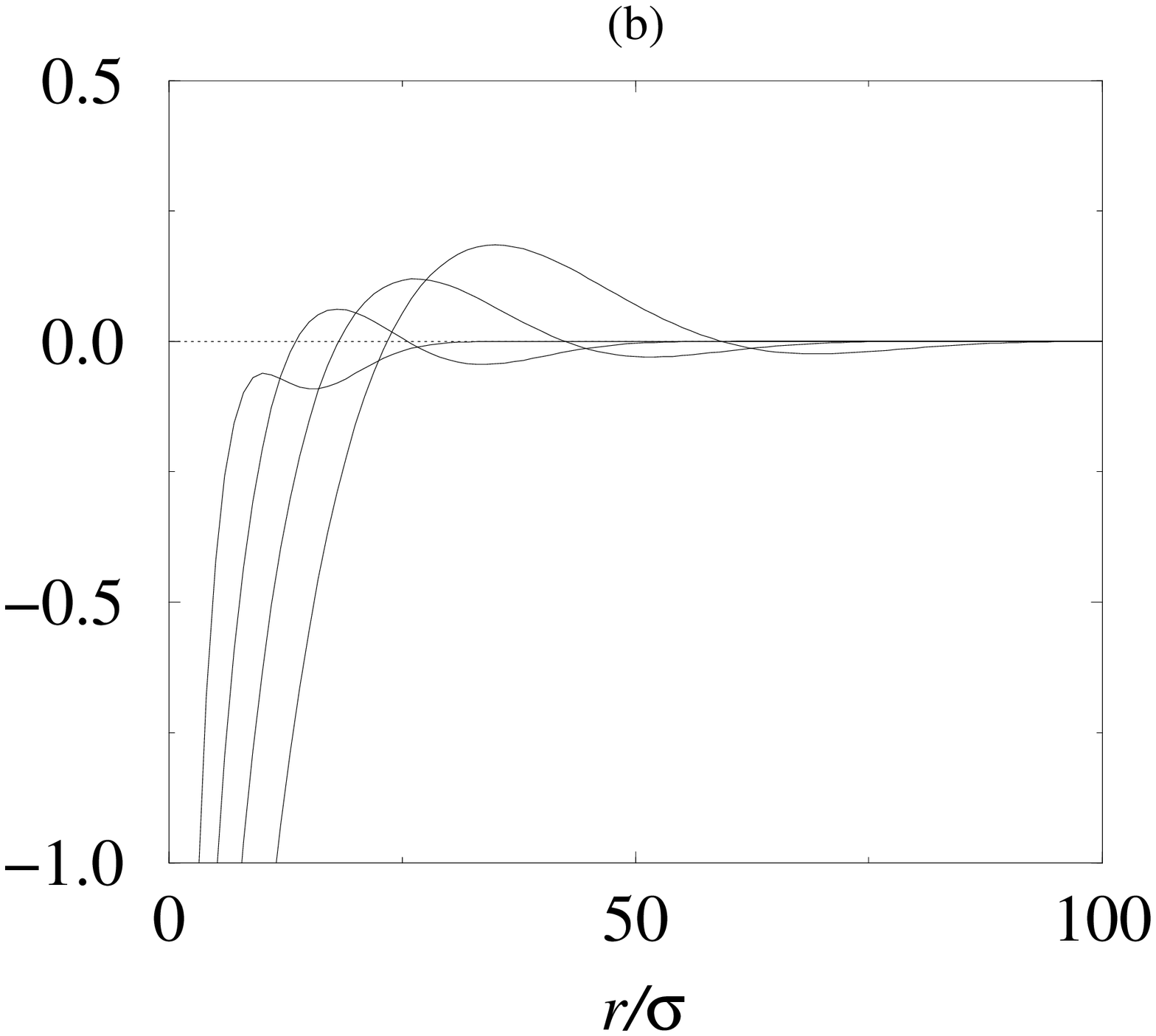,height=7.5cm}}
\vspace{1cm}
\caption{\label{fig:gab}}
\end{figure}
\end{center}

Fig.\ \ref{fig:gab} shows the spatial correlation functions
$G_{nn}$ and $G_{nT}$, which are the inverse Fourier
transforms of the structure factors $S_{ab}(k,t)$, shown in Fig.\
\ref{fig:sab}. 
The spatial density correlation
$G_{nn}(r,t)$, obtained numerically from $S_{nn}(k,t)$,
exhibits a negative correlation centered around a distance
which for large times grows as $\sqrt{\tau}$.
Comparison with simulation results confirms that the present 
theory correctly predicts the buildup of density correlations
in the linear time regime $\tau< \tau_c$.
For a more comprehensive comparison of the velocity and density
correlation functions with MD simulations at different densities
and different inelasticities, we refer to Ref.\
\cite{orza+brito+ernst}, where also the range of validity of the
present theories will be investigated.

In summary, the typical length scales, of the vortices 
$L_v(t)\sim \sqrt{\tau}$ and of the density clusters
$L_{\rm cl}(t)\sim \sqrt{\tau/\eps}$ in Sec.\ \ref{sec:devo},
also correspond to the typical length scales on which the
equal-time correlation functions $G_\perp(r,t)$, 
$G_\parallel(r,t)$ and $G_{nn}(r,t)$ show structure.
In zeroth approximation the HCS is an adiabatically cooling
equilibrium state with short range correlation functions, given in
(\ref{eq:delta}).
As time increases, long range correlations develop which are at
most of ${\cal O}(\eps^{d/2})$ [see (\ref{eq:glambda}) and
(\ref{eq:expl})], and extend far beyond the microscopic and
kinetic scales. 

\section{Conclusion}
\label{sec:concl2}
In this paper the structure factors
$S_{ab}(\bk,t)$
and corresponding spatial correlation
functions $G_{ab}(\br,t)$ have been
calculated and compared with 2D molecular dynamics simulations
for weakly inelastic hard disk systems,
where $\eps=1-\alpha^2$ is small.
For strongly inelastic systems the macroscopic equations are not
known.
For weakly inelastic systems we have assumed the hydrodynamic
equations of an elastic fluid, supplied with an energy sink
representing the collisional dissipation.
Also, we have assumed that the homogeneous cooling state (HCS) is an
adiabatically cooling equilibrium state, which is only correct to
lowest order in $\eps$.
In fact, in Ref.\ \cite{granmat,brilliantov+poeschel} 
the Enskog-Boltzmann
equation is used to calculate $\eps$-dependent corrections to the
velocity distribution in the HCS.
As follows from the Enskog-Boltzmann equation,
the pressure for inelastic hard
spheres decreases linearly with decreasing $\alpha$.
However, in our lowest order theory
it has been set equal to the pressure of elastic hard
spheres (EHS). 
Moreover, the collision frequency $\omega$ and the transport
coefficients of viscosity and heat conductivity have been assumed to
be given by the Enskog theory for EHS, but there are substantial
corrections for inelastic hard spheres.
Moreover, there appear new transport
coefficients, which are absent in EHS fluids \cite{sela,brey3}.

The most important parts of this paper are Secs.\
\ref{sec:meso} and \ref{sec:devo}, where the basic theory is
developed, modeled on the Cahn-Hilliard theory for spinodal
decomposition.
Based on the idea that the HCS for weakly
inelastic hard spheres is essentially an adiabatic equilibrium
state with a slowly changing temperature, we have formulated a
fluctuation-dissipation theorem for this state, which enabled us
to construct Langevin equations for hydrodynamic fluctuations.
The full set of coupled equations (\ref{eq:matrixfl}) for the structure
factors can only be solved numerically.
To understand the physical excitations that drive the
instabilities, we have presented a theoretical analysis of the
structure factors using
a spectral analysis of the unstable
Fourier modes. 
The structure factor $S_\perp(k,t)$ for the transverse flow
field can be calculated exactly.
For the longitudinal one,
$S_\parallel(k,t)$, we have obtained the simple analytic approximation
(\ref{eq:spar2}).
It only accounts for the dominant
contribution of the heat mode, and gives an almost quantitatively
correct description of $S_\parallel(k,t)$ for all times, but it is
missing the little dip which appears both in the numerical
results of Fig.\ \ref{fig:stl}(a), as well as in the MD
simulation results of Fig.\ \ref{fig:stl}(b).

The dynamics of the {\em transverse} and {\em longitudinal} flow
fields on the largest length scales are controlled by stable
purely diffusive modes with very different diffusivities.
There are no sound modes on these length scales.
The structure factors $S_\perp(k,t)$ and $S_\parallel(k,t)$ are
bounded at all $k$ and $t$ by their initial values, and they
develop spatial structure on length scales $L_v(t)\sim
2\pi\xi_\perp\sqrt{2\gamma_0\tau}$ and $L_\parallel(t)\sim
2\pi\xi_\parallel\sqrt{2\gamma_0\tau}$ respectively.

Agreement between the predictions of fluctuating hydrodynamics
{\em with} Langevin noise for $S_\parallel$ and $S_\perp$, and
the results of MD simulations is very good.

Calculation of the structure factor $S_{nn}(k,t)$ for the
density fluctuations is essentially a {\em linear} stability
ana\-ly\-sis, which describes the {\em early} stages of clustering.
It is expected to break down, because the density fluctuations are predicted to
grow at an exponential rate.
This is indeed seen to happen 
as the time $\tau$ approaches the crossover time
$\tau_c$ to the nonlinear clustering regime, as defined in the
caption of
Fig.\ \ref{fig:cooling}.
The agreement with MD simulations for $\tau < \tau_c$ is
again quite good.
Moreover, the simple analytic long wavelength approximation
(\ref{eq:snna}) agrees in the most relevant small-$k$ range
almost quantitatively with the numerical results of the theory
either with or without Langevin noise. 

One would expect that the agreement between theory and
simulations would also break down for $S_\parallel$ and
$S_\perp$ when $\tau$ approaches $\tau_c$.
However, it has been shown \cite{noije-prl,noije-prerc,orza+brito+ernst}
that $S_\parallel$ and $S_\perp$, calculated from this linear
theory, remain in agreement with the MD simulations for $\tau >
\tau_c$.
More surprisingly, the energy decay $E(t)$ in Fig.\
\ref{fig:cooling} has been quantitatively explained in Ref.\
\cite{brito+ernst} over the whole simulation interval ($\tau
\lesssim 150\simeq 2\tau_c$), using structure factors $S_\perp$ and
$S_\parallel$, calculated from the present linear theory.

The density structure factor $S_{nn}(k,t)$ shows that
spatial
density fluctuations in undriven IHS fluids are {\em unstable},
and lead to the formation of density clusters.
The linear instability is driven by longitudinal velocity
fluctuations and described by a coupling coefficient
of ${\cal O}(k^2)$ in $S_{nn}$.
The fluctuations in the flow field are only {\em relatively}
unstable, and do not lead to exponential growth of the
corresponding structure factors.
Nevertheless, the dissipative IHS fluid develops structure on
intermediate scales with typical length scales
$L_{\rm cl}(t)\sim \xi_\parallel\sqrt{\eps \tau}\sim
\sqrt{\tau/\eps}$ for the mean cluster sizes, and
$L_v(t)\sim \xi_\perp\sqrt{\eps\tau}\sim \sqrt{\tau}$ for the
mean vortex diameters.

In the literature \cite{goldhirsch}
the clustering instability has also been studied on the basis of
the nonlinear hydrodynamic Eqs.\ (\ref{eq:hydrofree}), which
show
that clustering in the {\em late} stages of evolution is driven
by the viscous heating term, $\eta |\bnabla {\bf u}|^2$.
The typical length scale of the clusters is then related to the
shear mode, $\xi_\perp=\sqrt{\nu/\omega\gamma_0}$, where $\nu$
is
the kinematic viscosity and $\omega$ the collision frequency.

Section \ref{sec:spatial2} deals with spatial correlations.
The assumption of incompressible flow for nearly elastic hard
spheres ($\alpha \gtrsim 0.9$) --- i.e.\ based on the
relative instability of shear modes only --- leads to spatial
velocity correlations, including algebraic $r^{-d}$ tails,
that are correct up to large distances ($r\lesssim 2\pi
\xi_\parallel$).
We have shown by explicit calculation that at
{\em small} inelasticities
$S_\parallel^+(k,t)$ essentially vanishes for all wave numbers
except at very small $k$ values ($k\lesssim 1/\xi_\parallel$),
where the assumption of incompressible
$\bf u$ fluctuations, made in subsection V.A,
breaks down.
Consequently, at small inelasticities
the most important qualitative modification that 
$S_\parallel^+$ adds to the spatial correlation function
$G_\parallel(r,t)$ of subsection V.A
is to provide an exponential cutoff for the $r^{-d}$ tail at the
largest scales $r \gtrsim 2\pi \xi_\parallel$.
At {\em larger} inelasticities the nonvanishing contributions from
$S^+_\parallel(k,t)$
modify $G_\parallel(r,t)$ and $G_\perp(r,t)$ significantly at all
distances.

The good quantitative correspondence between theory and
computer simulations shows that 
our theory for structure factors,
$S_{\alpha\beta}(\bk,t)$ and $S_{nn}(k,t)$, and spatial
correlation functions, $G_{\alpha\beta}(\br,t)$ and
$G_{nn}(r,t)$,
is correct for wave number, position and time
dependence in the relevant hydrodynamic range and for
inelasticities
$(\alpha \gtrsim 0.6)$ that are not too large.

Moreover, we have emphasized that there exist two different
mechanisms for structure formation and pattern selection, active
in granular fluids.
The first one, driving the formation of vortex structures, is
the mechanism of noise reduction \cite{brito+ernst}, 
comparable to peneplanation in structural geology with Ayers Rock
(Mount Uluru) as a spectacular example.
This mechanism
{\em selectively} suppresses velocity fluctuations at {\em shorter}
wavelengths as compared to longer ones, as well as {\em
longitudinal} fluctuations as compared to transverse ones.
The second mechanism, driving density clustering, is the more
common instability in density or composition fluctuations
as occurring in spinodal decomposition.
In freely evolving granular fluids the density is
most unstable at typical wavelengths $L_{\rm cl}(t)\sim 2\pi/k_{\rm
max}(t)$. 

\section*{Acknowledgments}
The authors want to thank R. Brito and J.A.G. Orza for many
stimulating
discussions and correspondence, and for the pleasant collaboration
in this joint project.
Also thanks are due to W. van de Water for
pointing out that relation (\ref{eq:incomp}) is well-known in the
theory of homogeneous turbulence.
M.E. wants to thank E. Ernst for providing the Ayers Rock analog,
and 
T.v.N. acknowledges support of the
foundation `Fundamenteel Onderzoek der Materie (FOM)', which is
financially supported by the Dutch National Science Foundation
(NWO).

{\appendix
\section{Appendix: transport coefficients}
\label{sec:app2a}
Linearization of the hydrodynamic equations (\ref{eq:hydrofree})
around the HCS results in the following set of equations:
\ba
\partial_t \delta n &=& - n \bnabla\cdot{\bf u}\nonumber\\
\partial_t {\bf u} &=& - \frac{1}{\rho}\bnabla p
+\nu \nabla^2 {\bf u}+(\nu_l-\nu) \bnabla \bnabla\cdot{\bf
u}\nonumber\\
\partial_t \delta T &=& \frac{2\kappa}{dn}\nabla^2\delta T
-\frac{2 p}{d
n}\bnabla\cdot{\bf u}-\delta \Gamma.
\label{eq:app2_1}
\ea
The pressure $p$ is assumed to be that of elastic hard spheres
(EHS), $p=n
T(1+\Omega_d\chi n
\sigma^d /2 d)$, where $\Omega_d=2 \pi^{d/2}/\Gamma(d/2)$ is the
$d$-dimensional solid angle, and $\chi(n)$ the equilibrium value
of
the pair correlation function of EHS of diameter $\sigma$ and
mass $m$ at
contact.
The kinematic and longitudinal viscosities $\nu$ and $\nu_l$, as
well
as the heat conductivity $\kappa$ are also assumed to be
approximately
equal to the corresponding quantities for EHS, as calculated from
the
Enskog theory \cite{chapman}, where $\rho\nu=\eta$ and $\rho
\nu_l=2\eta(d-1)/d+\zeta$ are expressed in shear viscosity $\eta$
and
bulk viscosity $\zeta$.
The collisional energy loss in (\ref{eq:Gloss}),
$\Gamma=2\gamma_0\omega T$, is proportional to the collision
frequency
\be
\omega =\Omega_d \chi n \sigma^{d-1}\sqrt{\frac{T}{\pi m}},
\label{eq:app2}
\ee
as obtained from the Enskog theory.
In three dimensions we use the Carnahan-Starling approximation
$\chi=(2-\phi)/2(1-\phi)^3$ \cite{carnahan}, in two dimensions the
Verlet-Levesque
approximation $\chi=(1-7\phi/16)/(1-\phi)^2$ \cite{verlet}.

In the body of the paper we need the eigenvalues $\zeta_\lambda(\bk)$
of the asymmetric matrix $\widetilde{M}_{ab}(\bk)$, introduced in 
(\ref{eq:lin2}), and its
right and left eigenvectors, which are obtained from
\ba
\widetilde{\bf M}(\bk)\tilde{\bw}_\lambda(\bk)&=&\zeta_\lambda(\bk)
\tilde{\bw}_\lambda(\bk)\nonumber\\
\widetilde{\bf M}^T(\bk)\tilde{\bv}_\lambda(\bk)&=&\zeta_\lambda(\bk) 
\tilde{\bv}_\lambda(\bk),
\label{eq:2_ev}
\ea
where $\widetilde{\bf M}^T$ is the transpose of $\widetilde{\bf M}$.
Here
$\lambda=\pm$
labels the sound modes, $\lambda=H$ the heat mode, and
$\lambda=\perp$
labels
$(d-1)$ degenerate shear or transverse velocity modes.
The eigenvectors form a complete biorthonormal basis, i.e.\
\ba
\sum_a \tilde{v}_{\lambda a}(\bk) \tilde{w}_{\mu a}(\bk)
&\equiv&\langle
\tilde{\bv}_\lambda|
\tilde{\bw}_\mu\rangle=\delta_{\lambda\mu}\nonumber\\
\sum_\lambda |\tilde{\bw}_\lambda(\bk)\rangle  \langle
\tilde{\bv}_\lambda(\bk)|&=&{\bf I},
\label{eq:2_compl}
\ea
where $\langle .|.\rangle$ is the usual scalar product.
The relations (\ref{eq:2_compl}) allow a spectral decomposition
of $\widetilde{\bf M}(\bk)$ in the form
\be
\widetilde{M}_{ab}(\bk)=\sum_\lambda \tilde{w}_{\lambda a}(\bk)
\zeta_\lambda(k) \tilde{v}_{\lambda b}(\bk).
\label{eq:spde}
\ee
Moreover the eigenvalue equation, ${\rm det}[\zeta(k){\bf I}-
\widetilde{\bf
M}(\bk)]=0$, is an {\em even} function of $\bk$.
Consequently, $\widetilde{\bf M}(\bk)$ and $\widetilde{\bf M}(-\bk)$ have 
the same
eigenvalues, which are either real or form a complex conjugate
pair.
So we choose $\zeta_\lambda(\bk)=\zeta_\lambda(-\bk)=
\zeta_\lambda(k)$.
The corresponding eigenvectors of $\widetilde{\bf M}(-\bk)$ in case of 
propagating sound
modes are obtained from the
transformation $\{\tilde{\bw}_+(-\bk),\tilde{\bv}_-
(-\bk)\}\leftrightarrow
\{\tilde{\bw}_{-}(\bk),\tilde{\bv}_{+}(\bk)\}$.
All other eigenvectors, corresponding to nonpropagating modes,
are invariant under the transformation
$\bk\rightarrow -\bk$.

The right and left eigenvectors of
the hydrodynamic matrix at
zero wave number
are given by
\ba
\tilde{\bw}_\perp(\bk)=(0,0,0,1)
\,\,\,\,\,\,\,\,\,\,\,\,\,\,\,\,\,\,\,\,\,\,
&&\tilde{\bv}_\perp(\bk)=(0,0,0,1)\nonumber\\
\tilde{\bw}_H(\bk)=(0,0,1,0)
\,\,\,\,\,\,\,\,\,\,\,\,\,\,\,\,\,\,\,\,\,\,
&&\tilde{\bv}_H(\bk)=( 0,0,1,0)\nonumber\\
\tilde{\bw}_+(\bk)=(1,-g(n),0,0)
\,\,\,\,\,\,\,\,\,\,
&&\tilde{\bv}_+(\bk)=
(1,0,0,0)\nonumber\\
\tilde{\bw}_-(\bk)=(0,1,0,0)
\,\,\,\,\,\,\,\,\,\,\,\,\,\,\,\,\,\,\,\,\,\,
&&\tilde{\bv}_-(\bk)=
(g(n),1,0,0).
\label{eq:eigenve2}
\ea
They characterize the $(d-1)$ shear modes ($\lambda=\perp$),
the heat mode ($\lambda=H$) and the two sound modes
($\lambda=\pm$).
Higher-$k$ corrections are given in Ref.\ \cite{brito+ernst}.
The shear or transverse velocity mode
is decoupled from the other modes, and
has a growth rate
$\zeta_\perp(k)=\gamma_0(1-k^2 \xi_\perp^2)$.

In the dissipative range ($k l_0\ll \gamma_0$), as
$k\rightarrow 0$,
the heat mode is a pure longitudinal
velocity
fluctuation, while the sound modes are a mixture of density and
temperature fluctuations.
To first order in $k$, density and temperature
fluctuations couple to the heat mode, and longitudinal velocity
fluctuations couple to the sound modes.
For large wave numbers ($k l_0\gg \gamma_0$)
the conventional character of heat and
sound
modes is recovered.
Besides the (uncoupled) shear mode, the
heat mode is unstable for $k<k_H^\ast$, where $k_H^\ast$ is the
root of $\zeta_H(k)=0$, or equivalently the root of ${\rm det}
[\widetilde{\bf M}(\bk)]=0$.
This yields the stability threshold, 
$k_H^\ast=[g(n)\chi_T p-1]^{1/2}/\xi_T\sim l_0/\sqrt{\eps}$.
Its dispersion
relation to second order in $k$ is given by
$\zeta_H(k)=\gamma_0(1-k^2
\xi_\parallel^2)$, where
\be
\xi_\parallel^2=\xi_l^2
+\frac{l_0^2}{2\gamma_0^2}\left[\frac{1}{n T
\chi_T}-\left(\frac{p}{n
T}\right)\left(1+\frac{n}{\chi}\frac{\partial
\chi}{\partial n} - \frac{p}{d n T}\right)\right].
\label{eq:xip}
\ee
Note that $\xi_\parallel$ diverges as $1/\eps$, while
$\xi_\perp\sim 1/\sqrt{\eps}$; as a consequence the correlation
lengths $\xi_\parallel$ and $\xi_\perp$ are well separated for
small inelasticity \cite{noije-prerc}.
 
The sound modes are stable for all wave numbers.
In the dissipative regime ($k l_0\ll \gamma_0$), they correspond
to nonpropagating modes.
The mode labeled $\lambda=+$ is for small $k$ a linear combination of 
a density and a temperature 
fluctuation with $\zeta_+(k)\simeq -\gamma_0 k^2 \xi_+^2$, and
the mode labeled $\lambda=-$ is a pure temperature fluctuation,
with $\zeta_-(k)\simeq \gamma_0(-1+k^2 \xi_-^2)$, which is a
kinetic mode.
To lowest order in $k$ the new correlation lengths are given by
\ba
\xi_+^2(k)&=&\frac{l_0^2}{2 \gamma_0^2 }\left[2
\left(\frac{p}{nT}\right)
\left(1+\frac{n}{\chi}\frac{\partial
\chi}{\partial n}\right)-\frac{1}{n T
\chi_T}\right]\nonumber\\
\xi_-^2(k)&=& \left[-\xi_T^2+\frac{l_0^2}{2
\gamma_0^2}
\left(\frac{p}{n T}
\right)\left(1+\frac{n}{\chi}\frac{\partial
\chi}{\partial n}+\frac{p}{d n T}\right)\right].
\label{eq:z+-}
\ea
At a wave number of the order of $\gamma_0/l_0$,
their dispersion relations merge and the modes
start to propagate.

\section{Appendix: Fourier transforms}
\label{sec:app2b}
To calculate the tensor velocity correlation function
$G_{\alpha\beta}(\br,t)$ by Fourier inversion from
$S_{\alpha\beta}(\bk,t)$, we start from Eqs.\ (\ref{eq:sdec}),
(\ref{eq:gdec}), (\ref{eq:gdel}) and (\ref{eq:splus}), and
consider first the {\em incompressible} limit where
$S_\parallel(k,t)=0$, i.e.\
\be
G_{\alpha\beta}^+(\br,t)=\int \frac{{\rm d}\bk}{(2\pi)^d} 
\exp(i\bk\cdot \br)
(\delta_{\alpha\beta}-\hat{k}_\alpha\hat{k}_\beta)S_\perp^+(k,t).
\ee
According to (\ref{eq:gdec}) $G_{\alpha\beta}$ can be split into
two scalar functions, $G_\parallel$ and $G_\perp$, which will be
expressed in $S_\perp$.
The simplest functions to calculate are the trace and
parallel part of $G_{\alpha\beta}$, i.e.\
\ba
G^+_{pp}(r,t)&=&\sum_\alpha G^+_{\alpha\alpha}(\br,t)\nonumber\\
&=& (d-1) \int
\frac{{\rm
d}\bk}{(2\pi)^d} \exp(i\bk\cdot\br) S^+_\perp(k,t)\nonumber\\
G^+_\parallel(r,t)&=& \hat{r}_\alpha\hat{r}_\beta
G^+_{\alpha\beta}(\br,t)\nonumber\\
&=&\int
\frac{{\rm
d}\bk}{(2\pi)^d} \exp(i\bk\cdot\br) [1- (\hat{\bk}\cdot
\hat{\br})^2] S^+_\perp(k,t).\nonumber\\
\label{eq:begin}
\ea
To carry out the $d$-dimensional angular integrations for $d\geq
2$ we express the infinitesimal solid angle as
\be
{\rm d}\hat{\bk}=(\sin{\theta_1})^{d-2} \dots
(\sin{\theta_{d-2}}) {\rm d}\theta_1 \dots {\rm
d}\theta_{d-2}{\rm d}\phi,
\label{eq:angi}
\ee
where $\theta_n \in (0,\pi)$ are polar angles and $\phi \in
(0,2\pi)$ is an azimuthal angle, and we note that the full solid
angle is $\Omega_d=\int {\rm d}\hat{\bk} =
2\pi^{d/2}/\Gamma(d/2)$.
Then we use the relation
\ba
\int \frac{{\rm d}\hat{\bk}}{\Omega_d} \exp(i\bk\cdot\br) &=&\frac{
\int_0^\pi {\rm d}\theta (\sin{\theta})^{d-2} \exp(i k r
\cos{\theta})}
{\int_0^\pi {\rm d}\theta (\sin{\theta})^{d-2}}\nonumber\\
&=&\left(\frac{2}{kr}\right)^{d/2-1}\Gamma(d/2) J_{d/2-1}(kr),
\label{eq:bes}
\ea
where the integral representation (8.411.7) of Ref.\
\cite{gradshteyn} has been used for the Bessel function
$J_\nu(z)$.
Then Eqs.\ (\ref{eq:begin}) become
\ba
G^+_{pp}(r,t)&=&\frac{d-1}{(2\pi)^{d/2}r^{d/2-1}}
\int_0^\infty
{\rm d}k k^{d/2} J_{d/2-1}(kr) S^+_\perp(k,t)\nonumber\\
G^+_\parallel(r,t)&=&\frac{d-1}{(2\pi)^{d/2}r^{d/2}}\int_0^\infty
{\rm d}k k^{d/2-1}J_{d/2}(kr) S^+_\perp(k,t).\nonumber\\
\label{eq:tracepar}
\ea
With the help of the recursion formula for Bessel functions,
$z {\rm d}J_\nu(z)/{\rm d}z+\nu J_\nu(z)=z J_{\nu-1}(z)$,
together with the general relation
\be
G^+_\perp(r,t)=[G^+_{pp}(r,t)-G^+_\parallel(r,t)]/(d-1),
\label{eq:gperp}
\ee
we obtain $G^+_\perp(r,t)$ from
$G^+_\parallel(r,t)$, i.e.\
\be
G^+_\perp(r,t)=G^+_\parallel(r,t)+\left(\frac{r}{d-1}\right)
\frac{\partial}{\partial r} G^+_\parallel(r,t).
\label{eq:incomp}
\ee
This is a well known relation in the theory of homogeneous and 
isotropic
turbulence in 
incompressible flows (see Refs. \cite{batchelor} and
\cite{landau}, Chap.\ 3).

In the general case $S_\parallel(k,t)$ is nonvanishing and we
have from (\ref{eq:sdec}) an additional part, denoted by
$\overline{G}_{\alpha\beta}(r,t)$, coming from $\hat{k}_\alpha
\hat{k}_\beta S_\parallel(k,t)$.
Here we have the relations
\ba
\overline{G}^+_{pp}(r,t)&=& 
\int
\frac{{\rm
d}\bk}{(2\pi)^d} \exp(i\bk\cdot\br) S^+_\parallel(k,t)\nonumber\\
\overline{G}^+_\perp(r,t)&=& 
 \frac{1}{d-1}\int
\frac{{\rm
d}\bk}{(2\pi)^d} \exp(i\bk\cdot\br) [1- (\hat{\bk}\cdot
\hat{\br})^2] S^+_\parallel(k,t).\nonumber\\
\label{eq:begin1}
\ea
The results for these functions can be read off directly from
(\ref{eq:begin}) and (\ref{eq:tracepar}), and the parallel
part is obtained from (\ref{eq:begin1}) as
\ba
\overline{G}_\parallel^+(r,t)&=&\overline{G}_{pp}^+(r,t)-
(d-1)\overline{G}_\perp^+(r,t)\nonumber\\
&=&\overline{G}_{\perp}^+(r,t)+
r \frac{\partial}{\partial r} \overline{G}^+_\perp(r,t).
\ea
The derivation of the second line parallels that of
Eq.\ (\ref{eq:incomp}). Note also that the role of
$\overline{G}_\parallel$ and $\overline{G}_\perp$ has been
interchanged with respect to the incompressible case.

Fourier inversion of any of the {\em scalar} functions
$S_{ab}(k,t)$ with $a,b=\{n,T\}$ is covered by the first line of
Eqs.\ (\ref{eq:begin}) and (\ref{eq:tracepar}).
We consider first $G_\parallel$ and $G_\perp$ in the
{\em incompressible limit}, where $S_\perp(k,t)$ is given by
(\ref{eq:sperp2}).
The Cahn-Hilliard theory in the incompressible limit will be
considered later.

Inspection of (\ref{eq:sperp2}) shows that the large-$k$ limit
of $S_\perp(k,t)$ is $S_\perp^{\infty}=T(t)/mn$, leaving
$S_\perp^+(k,t)$ as a remainder.
This may be written as
\be
S_\perp^+(k,t)=\frac{T(t)}{mn} \int_0^{2\gamma_0\tau} {\rm
d}s^\prime \exp[(1-k^2\xi_\perp^2)s^\prime].
\label{eq:sp}
\ee
Using Eq.\ (\ref{eq:tracepar}) for the parallel velocity
correlation function $G_\parallel^+(r,t)$ and (\ref{eq:incomp})
to determine $G_\perp^+(r,t)$, we obtain 
$G_\lambda^+(r,t)=(T(t)/mn\xi_\perp^d) g_\lambda(x,s)$ for
$\lambda=\{\parallel,\perp\}$ with $s=2\gamma_0\tau$ and 
$x=r/\xi_\perp$, which is is valid in dimensions $d\geq 2$.
Moreover, $g_\lambda(x,s)$ is given by
\ba
g_\parallel(x,s)&=&
\frac{d-1}{2\pi^{d/2}x^d}\int_0^s
{\rm
d}s^\prime
\exp(s^\prime)\gamma{\left(\frac{d}{2},\frac{x^2}{4s^\prime}\right)}
\nonumber\\
g_\perp(x,s)&=&\int_0^s{\rm
d}s^\prime\exp(s^\prime)
\frac{\exp(-
x^2/4s^\prime)}{(4 \pi s^\prime)^{d/2}}
-\frac{1}{d-1}g_\parallel(x,s).\nonumber\\
\label{eq:glamb}
\ea
The Bessel transform (\ref{eq:tracepar}) of $\exp(-\beta k^2)$ in
(\ref{eq:sp}) has been calculated using Eq.\ (6.631.5) of Ref.\
\cite{gradshteyn}, where $\gamma(\alpha,z)=\int_0^z {\rm d}t
\exp(-t) t^{\alpha-1}$ is the incomplete gamma function.
For $d=2$ it reduces to $\gamma(1,z)=1-\exp(-z)$ and for $d=3$
to $\gamma(3/2,z)=\sqrt{\pi}\phi(\sqrt{z})/2-\sqrt{z}\exp(-z)$,
where $\phi(z)$ is the error function.
We observe that $g_\parallel(x,s)$ is {\em positive} for all
$x,s$. 
For large distance, $x^2\gg 4s$, the functions $g_\lambda(x,s)$
show algebraic tails $\sim 1/x^d$.
This can be seen by noting that $\gamma(\alpha,x^2/4s)$
approaches $\Gamma(\alpha)$, so that 
\be
g_\parallel(x_\perp,s)=-(d-1) g_\perp(x_\perp,s)\sim
\left(\frac{d-1}{\Omega_d x_\perp^d}\right)
[\exp(s)-1],
\label{eq:tail}
\ee
where $\Omega_d$ is defined below (\ref{eq:angi}).

The theory {\em without} noise for the transverse structure function
is according to the discussion in Sec.\ \ref{sec:devo} given by
\be
S_\perp(k,t)=S_\perp^+(k,t)=\frac{T(t)}{mn} \exp[2\gamma_0(1-k^2
\xi_\perp^2)\tau].
\label{eq:sperpch}
\ee
In the {\em incompressible} limit, where $S_\parallel(k,t)=0$,
the velocity correlations in the flow field can be calculated by
Fourier inversion of (\ref{eq:sperpch}).
Comparison of (\ref{eq:sperpch}) with (\ref{eq:sp}) shows that 
\be
G_\lambda(r,t)=\frac{T(t)}{mn\xi_\perp^d}
g_\lambda\left(\frac{r}{\xi_\perp},
2\gamma_0 \tau\right),
\ee
where
\ba
g_\parallel(x,s)&=&
\frac{\partial g_\parallel(x,s)}{\partial s}
= \frac{d-1}{2\pi^{d/2}x^d} \exp(s)
\gamma{\left(\frac{d}{2},\frac{x^2}{4s}\right)}\nonumber\\
g_\perp(x,s)&=&\exp(s)\frac{\exp(-
x^2/4s)}{(4\pi s)^{d/2}} -\frac{1}{d-1}g_\parallel
(x,s).
\ea

}
\end{multicols}
\newpage
\begin{center}
{FIGURE CAPTIONS}
\end{center}
 
\begin{center}
\begin{figure}[h]
\centerline{\hspace{-8.5cm}\psfig{figure=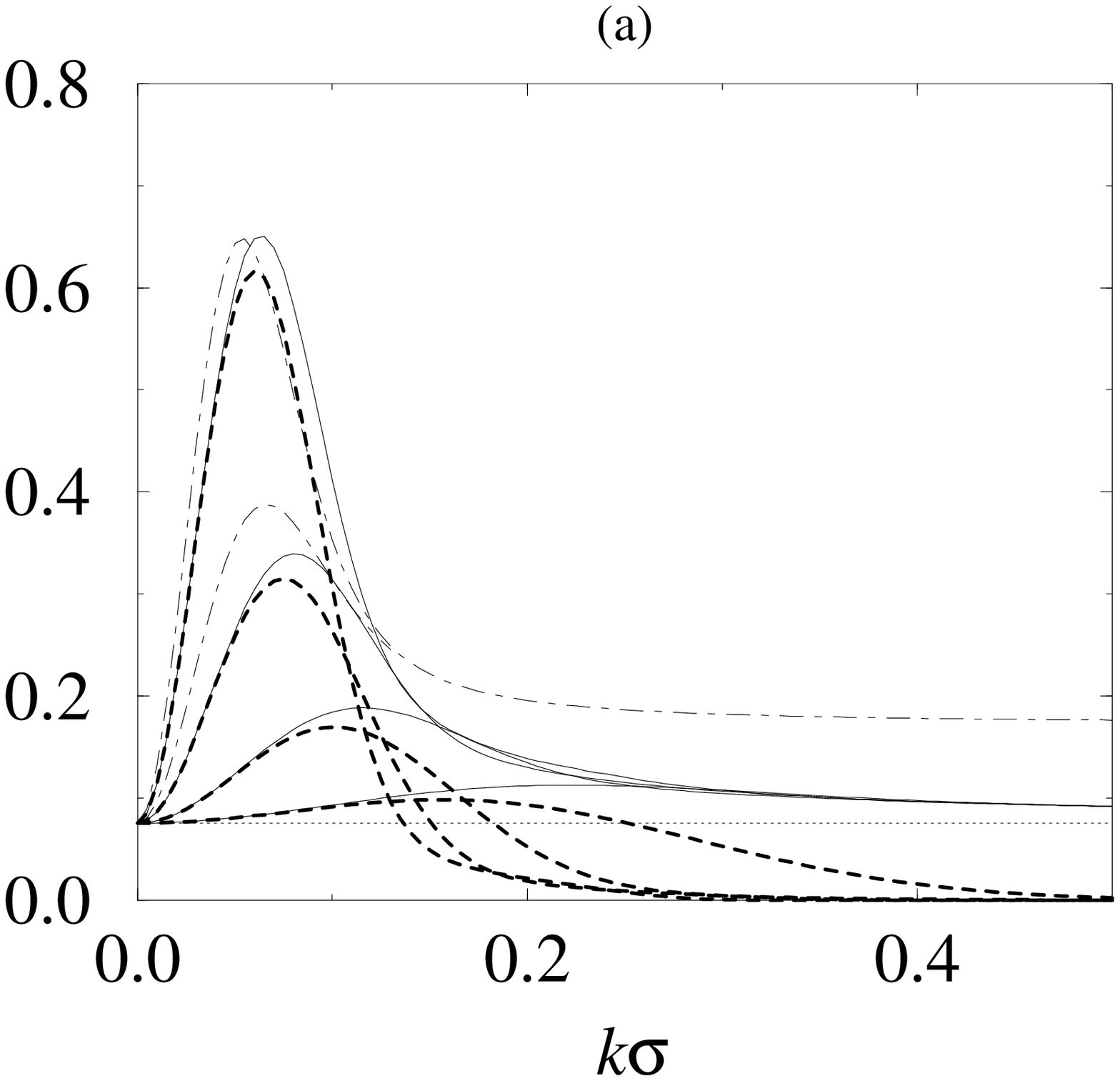,height=7.5cm
}}
\vspace{-7.5cm}
\centerline{\hspace{8cm}\psfig{figure=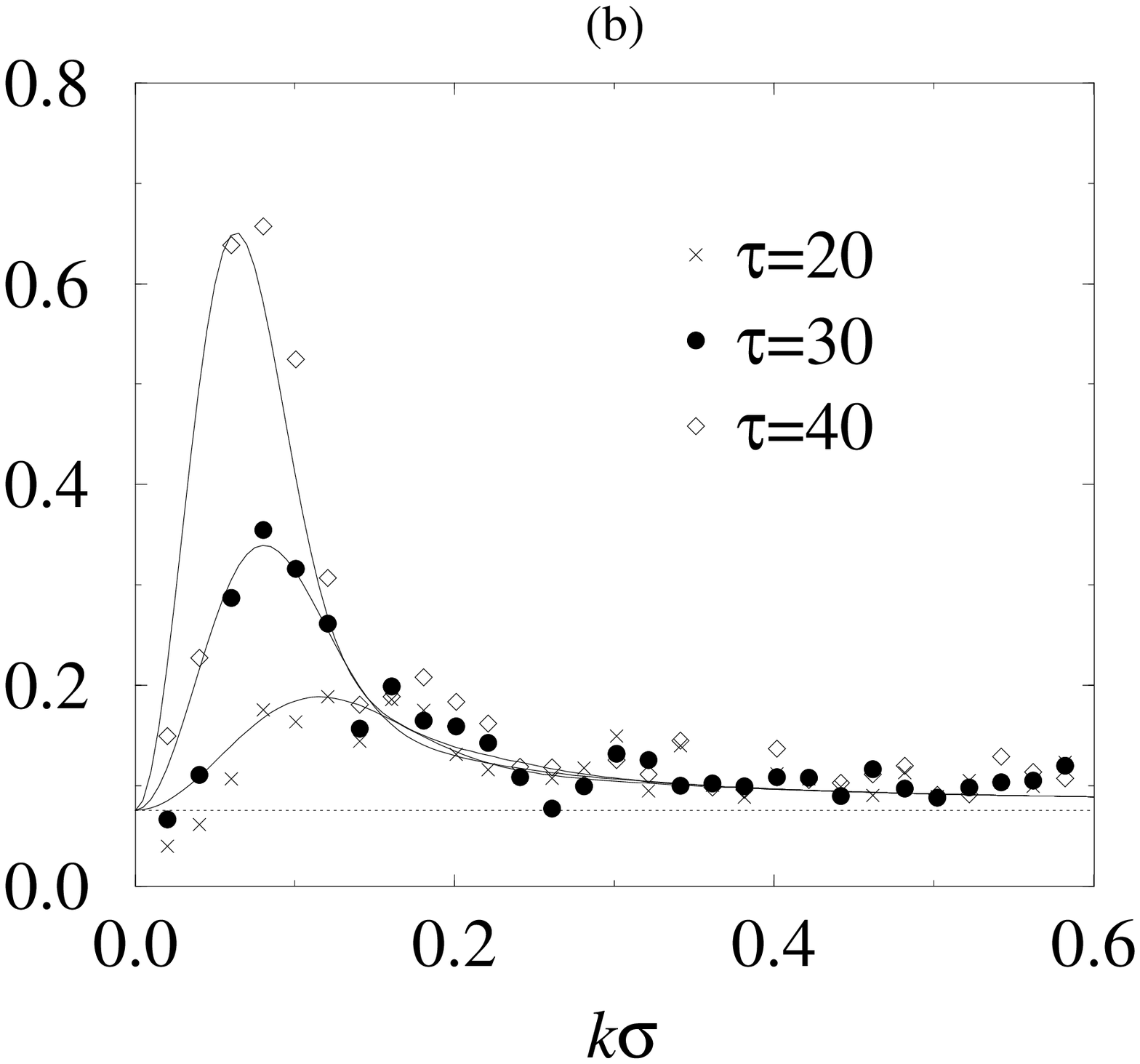,height=7.5cm}}
\vspace{1cm}
\caption{\label{fig:snn}}
\end{figure}
\end{center}

\begin{center}
\begin{figure}[h]
\centerline{\hspace{-8.5cm}\psfig{figure=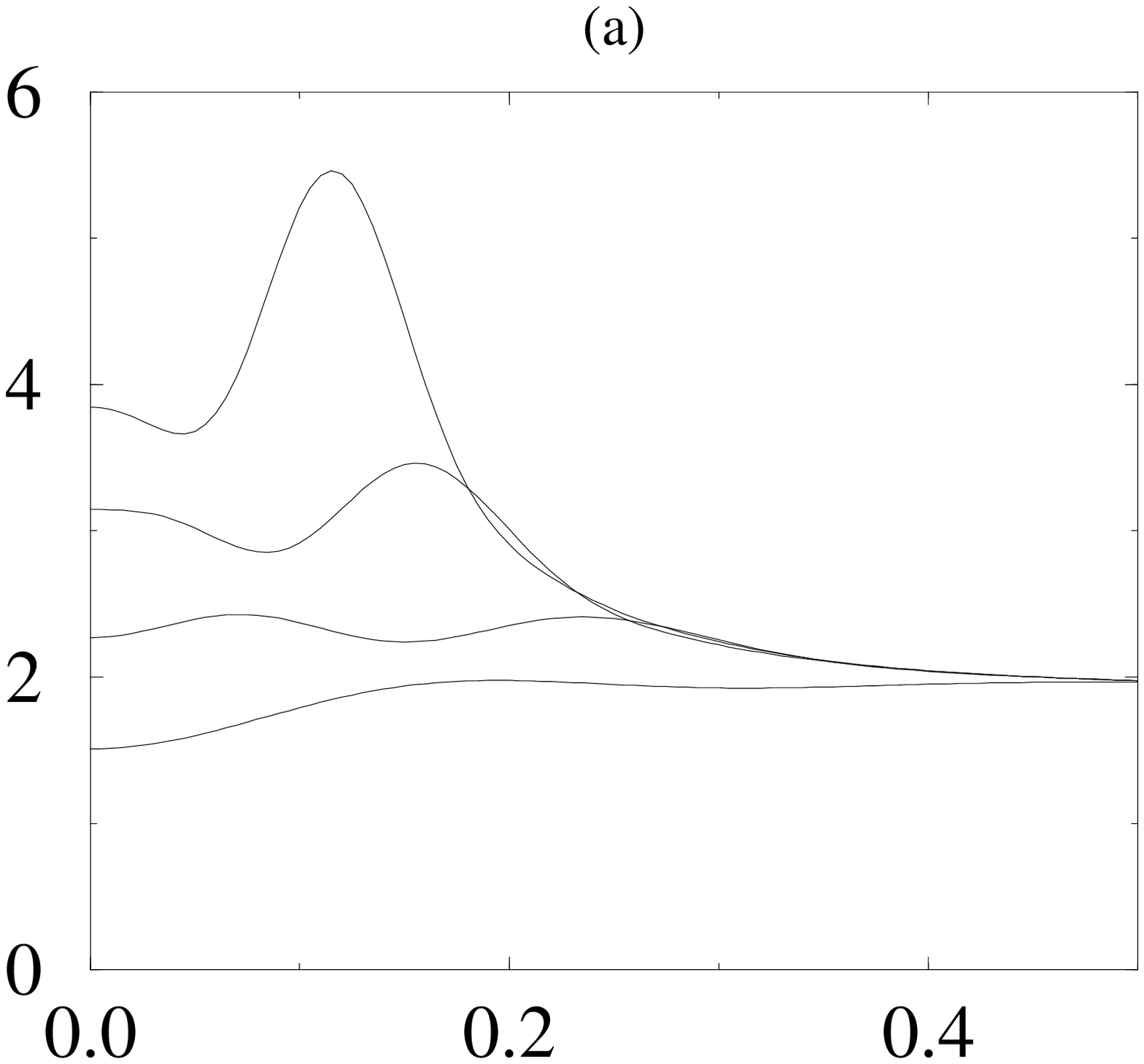,height=7.5cm}}
\vspace{-7.5cm}
\centerline{\hspace{8cm}\psfig{figure=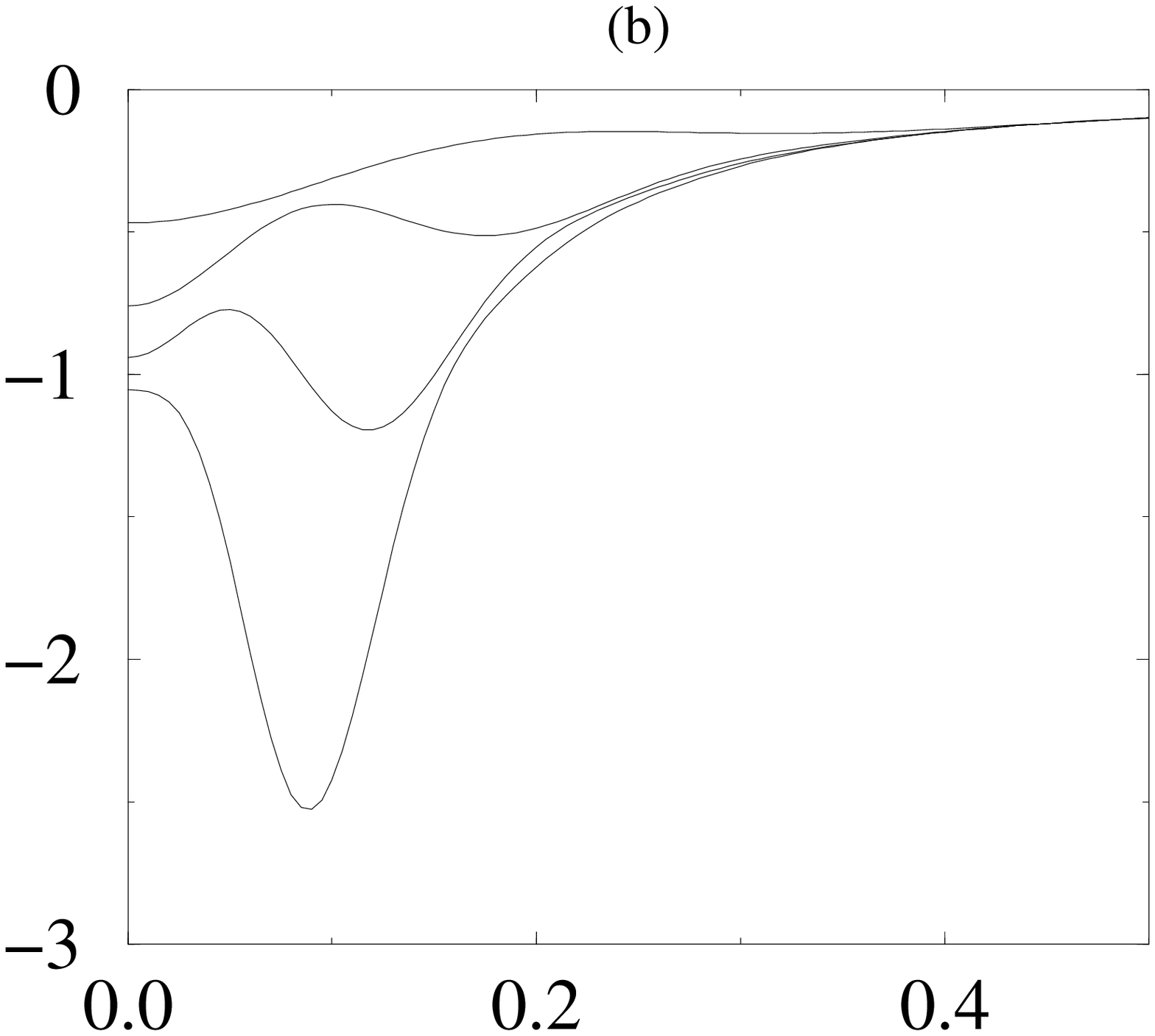,height=7.5cm}}
\centerline{\hspace{-8.5cm}\psfig{figure=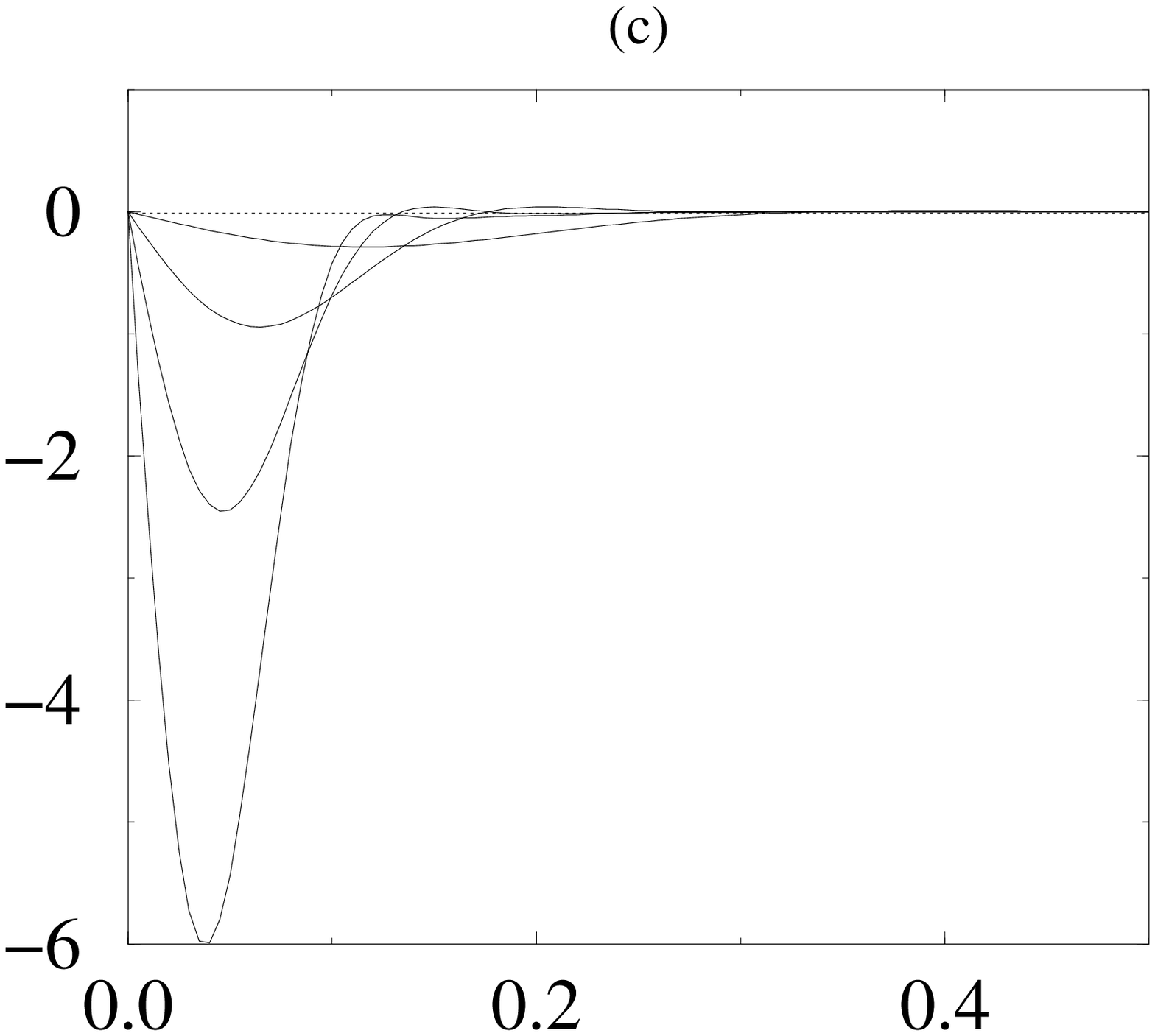,height=7.5cm}}
\vspace{-7.5cm}
\centerline{\hspace{8cm}\psfig{figure=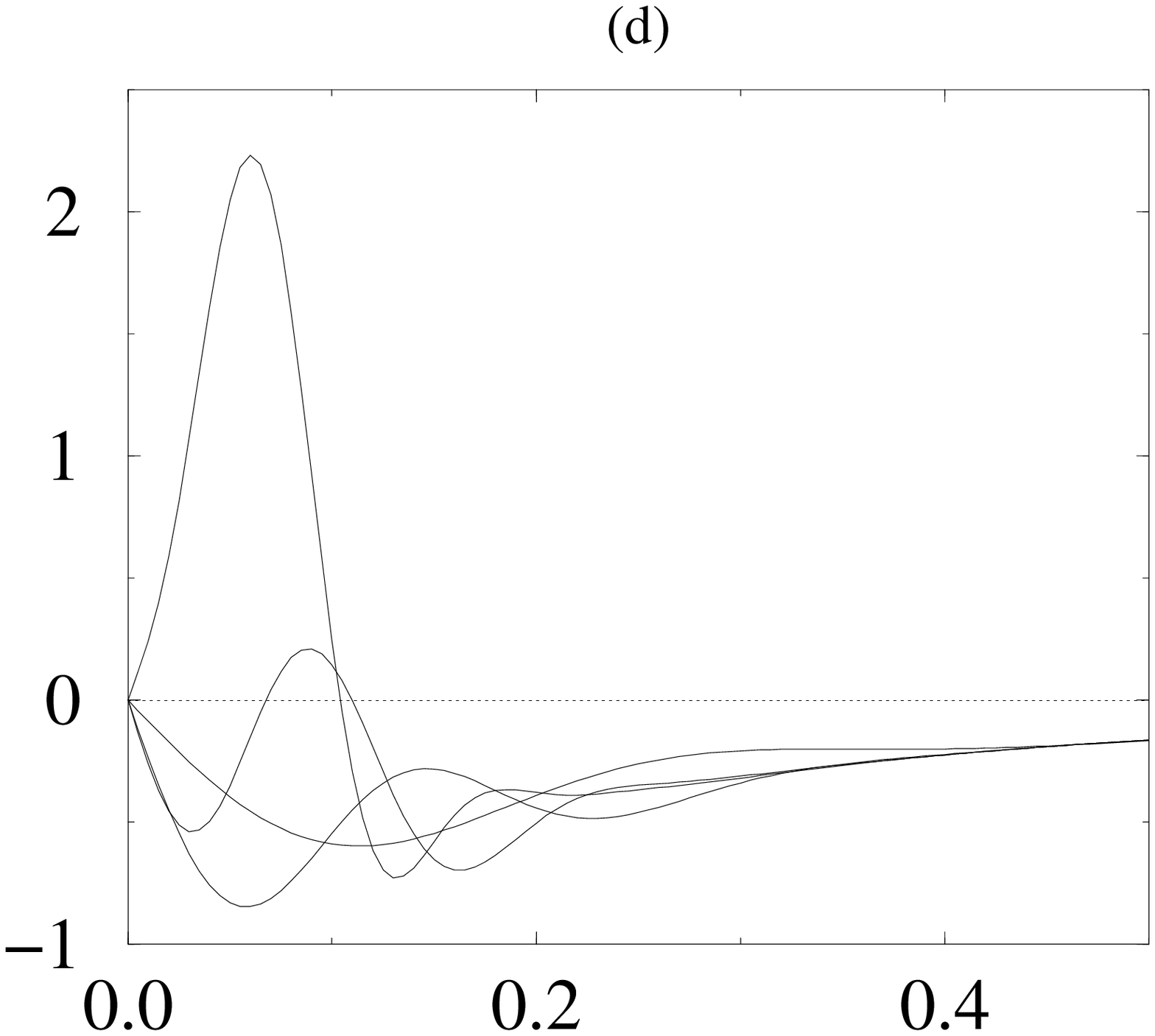,height=7.5cm}}
\vspace{1cm}
\caption{\label{fig:sab}}
\end{figure}
\end{center}

\begin{enumerate}
\item
Left: Velocity field after $\tau=80$ collisions per
particle. The density is then still more or less homogeneous.
Right: Density field at $\tau=160$. System of inelastic
hard disks at a packing fraction
$\phi=\textstyle{\frac{1}{4}}\pi\sigma^2=0.4$ and $\alpha=0.9$.

\item
Kinetic energy per particle $E$ versus number of
collisions per particle $\tau$ for
$\phi=0.4$  and $\alpha=0.9$.
Initially $E$ is equal to the temperature $T$ and follows Haff's
homogeneous cooling law (\ref{eq:haff}). The arrow indicates a
crossover time $\tau_c\simeq 70$ from the homogeneous
cooling state to the
nonlinear clustering regime.
Then spatial
inhomogeneities become important and slow down the energy decay.
The dashed line represents Haff's law (\ref{eq:haff})
and the dot-dashed line the result of Ref.\
\cite{brito+ernst} for the long time energy decay.

\item
Growth rates $\zeta_\lambda/\gamma_0$ for shear
($\lambda=\perp$), heat ($\lambda=H$) and sound ($\lambda=\pm$)
modes
versus $k \sigma$ for inelastic hard disks with $\alpha=0.9$ in
(a) and $\alpha=0.6$ in (b) at
a packing fraction $\phi=0.4$ ($l_0\simeq 0.34 \sigma$).
The shear and heat mode are unstable for $k<k_\perp^\ast$ and
$k<k_H^\ast$
respectively.
Imaginary parts of the sound modes, indicated by the dashed
lines, vanish for $k l_0\ll \gamma_0$.

\item
Ratio $\xi_\parallel/\xi_\perp$ versus packing fraction
$\phi$ of inelastic disks, with definition of $\xi_\parallel$ in
(\ref{eq:xip}) and of $\xi_\perp$ below (\ref{eq:change}).

\item
Density structure factor $S_{nn}$, in units $1/\sigma^2$,
versus
$k\sigma$ for $\phi=0.4$ and $\alpha=0.9$, at $\tau=10,20,30$ and 40
collisions per particle, exhibits the clustering instability with a
growing maximum at $k_{\rm max}(t)$, which shifts to the left.
Solid and dashed lines are the numerical solutions
of (\ref{eq:stab})
and (\ref{eq:stabch}) with and without Langevin noise
respectively.
They differ appreciably, except at small $k$.
The simple analytic approximation (\ref{eq:snna}),
shown in (a) as dot-dashed lines for $\tau=30$ and
40, gives a good description in the long time and long
wavelength limit. (b) Numerical solution of (\ref{eq:stab})
compared with MD simulation results
(courtesy J.A.G.\ Orza et al.\ \cite{orza+brito+ernst}).

\item
Structure factors of velocity fluctuations $S_\parallel$
and $S_\perp$,
in units
$T_0
\sigma^2/m$,
versus $k\sigma$ for $\phi=0.4$ and $\alpha=0.9$ illustrate the
phenomenon of noise reduction at small wavelengths.
(a) $S_\parallel$ at $\tau=10,20,30$ and 40, where solid lines
represent the numerical solution of (\ref{eq:stab}), and the
dot-dashed lines represent the approximate analytic result
(\ref{eq:spar2}).
The dashed line represents the numerical solution of the
`noiseless' Cahn-Hilliard theory (\ref{eq:stabch}) at $\tau=10$,
which deviates substantially from the solid line at $\tau=10$,
except near $k=0$.
(b) Comparison with MD simulation results (courtesy of J.A.G.\ Orza
et al.
\cite{orza+brito+ernst})
at $\tau=20$ for
$S_\perp$ (squares) and
$S_\parallel$ (circles).
The numerical solutions of (\ref{eq:stab}) and
(\ref{eq:stabch}) with (solid lines) and without Langevin noise
(dashed lines) approach each other for long times
and long wavelengths.

\item
Rescaled structure factors (a) $\tilde{S}_{TT}$, (b)
$\tilde{S}_{nT}$, (c) ${\rm Im}(\tilde{S}_{nl})$, and (d)
${\rm Im}(\tilde{S}_{Tl})$, all in units $\sigma^2$,
versus $k\sigma$ for
same parameters as used in Fig.\ \ref{fig:snn}.
Both $\tilde{S}_{nl}$ and $\tilde{S}_{Tl}$ change sign under
$\bk\rightarrow -\bk$ and
are therefore purely imaginary. The structure factors in (b),
(c) and (d) vanish initially and develop structure
as time increases. $\tilde{S}_{TT}$ develops structure on top of
its initial (plateau) value $2/dn$.

\item
Comparison of
theoretical predictions (\ref{eq:g2d}), based on
incompressibility,
with MD simulation results for $G_\parallel$ (filled circles) and
$G_\perp$ (open circles), in units $T_0/m$, versus $r/\sigma$
At $\phi=0.4$ and $\alpha=0.9$. 

\item
(a) $g_\parallel(x,s)$ and (b) $g_\perp(x,s)$ versus
$x=x_\perp=r/\xi_\perp$
for $s=2\gamma_0\tau=2$.
The solid lines corresponds to (\ref{eq:g2d}) in the
incompressible limit ($\xi_\parallel/\xi_\perp\rightarrow
\infty$), and the dashed lines to approximations
(\ref{eq:expl}),
for $\xi_\parallel/\xi_\perp=1,2,5,10$.
As $\xi_\parallel/\xi_\perp$ decreases, the $r^{-2}$ tail
in (a) is cut off exponentially at smaller distances and
finally disappears at $\xi_\parallel=\xi_\perp$.
The depth of the
minimum in (b) decreases with decreasing
$\xi_\parallel/\xi_\perp$
and finally disappears at
$\xi_\parallel=\xi_\perp$.

\item
Perpendicular velocity correlation $G_\perp$, in units
$T_0
/m$,
versus $r/\sigma$ for packing fraction
$\phi=0.4$, relatively high inelasticity
$\alpha=0.6$ and $\tau=40$.
Simulation results are compared with prediction (\ref{eq:g2d})
of the incompressible theory (dashed line), and the numerical
solution (solid line) of the full set of fluctuating hydrodynamic
equations.

\item
Spatial correlation functions (a) $G_{nn}(r,t)$, in units
$10^{-3}/\sigma^4$, and (b) $G_{nT}(r,t)/T(t)$, in units
$10^{-3}/\sigma^2$, versus $r/\sigma$ obtained
numerically from the structure factors shown in Fig.\
\ref{fig:sab} at the same parameters as used in Fig.\
\ref{fig:snn}.
Both functions show a growing correlation
length.

\end{enumerate}

\newpage

\newpage

\newpage

\newpage

\newpage

\newpage

\newpage

\newpage

\end{document}